\documentclass[preprint,aps,12pt,showpacs,nofootinbib,tightenlines,amsmath,amssymb]{revtex4}

\usepackage{graphicx}
\usepackage{bbm}
\usepackage{subfigure}

\newcommand{\be}{\begin{eqnarray}}
\newcommand{\beq}{\begin{eqnarray}}
\newcommand{\befg}{\begin{figure}}

\newcommand{\edfg}{\end{figure}}
\newcommand{\eeq}{\end{eqnarray}}
\newcommand{\ee}{\end{eqnarray}}

\newcommand{\PP}{{1 + v\hspace{-0.2cm}\slash \over 2}}
\newcommand{\PM}{{1 - v\hspace{-0.2cm}\slash \over 2}}

\newcommand{\vslash}{\not\!{v}}

\newcommand{\DL}{\overleftarrow D}
\newcommand{\DCLS}{\overleftarrow D^2_{\! \! \bot} }

    \newcommand{\DSC}{D\hspace{-0.25cm}\slash_{\bot}}
    \newcommand{\DSP}{D\hspace{-0.25cm}\slash_{\|}}
    \newcommand{\DS}{D\hspace{-0.25cm}\slash}
    \newcommand{\DC}{D_{\bot}}

    \newcommand{\QV}{Q_v^{+}}
    \newcommand{\QVB}{\bar{Q}_v^{+}}
    
    \newcommand{\QVBP}{\bar{Q}^{\prime +}_{v^{\prime}} }

    \newcommand{\QVH}{\hat{Q}_v}
    \newcommand{\QVHB}{\bar{\hat{Q}}_v}
    \newcommand{\VS}{v\hspace{-0.2cm}\slash}
    \newcommand{\MQ}{m_{Q}}
    
    \newcommand{\QVHPMB}{\bar{\hat{Q}}_v{\vspace{-0.3cm}\hspace{-0.2cm}{^{\pm}} }}
    
    \newcommand{\QVHPM}{\hat{Q}^{\pm}_v}
    \newcommand{\QVHMP}{\hat{Q}^{\mp}_v}

\makeatletter
\def\underbracket{%
    \@ifnextchar[{\@underbracket}{\@underbracket [\@bracketheight]}%
}
\def\@underbracket[#1]{%
    \@ifnextchar[{\@under@bracket[#1]}{\@under@bracket[#1][0.4em]}%
}
\def\@under@bracket[#1][#2]#3{%\message {Underbracket: #1,#2,#3}
    \mathop{\vtop{\m@th \ialign {##\crcr $\hfil \displaystyle {#3}\hfil $%
    \crcr \noalign {\kern 3\p@ \nointerlineskip }\upbracketfill {#1}{#2}
     \crcr \noalign {\kern 3\p@ }}}}\limits}
\def\upbracketfill#1#2{$\m@th \setbox \z@ \hbox {$\braceld$}
                  \edef\@bracketheight{\the\ht\z@}\bracketend{#1}{#2}
                  \leaders \vrule \@height #1 \@depth \z@ \hfill
                  \leaders \vrule \@height #1 \@depth \z@ \hfill \bracketend{#1}{#2}$}
\def\bracketend#1#2{\vrule height #2 width #1\relax}
\makeatother

\begin{document}

\title{Heavy Quark Expansion in $1/\hat{m}_Q$ and $|V_{cb}|$ Extraction}
\author{Wen-Yu Wang$^{1,2}$, \mbox{Yue-Liang Wu}$^{2}$, Fang Ye$^{2}$}
\affiliation{$^1$School of Applied Science,
University of Science and Technology Beijing, \\
 Beijing 100083, China \\
 $^2$Kavli Institute for Theoretical Physics China \\
 Key Laboratory of Frontiers in Theoretical Physics \\
 Institute of Theoretical Physics, Chinese Academy of Sciences,
P.O. Box 2735, Beijing 100190, China}

\begin{abstract}

The dressed heavy quark mass $\hat{m}_Q=m_Q+\bar{\Lambda}$ with
$\bar{\Lambda}$ being the binding energy is introduced to
characterize the heavy hadrons containing a single heavy quark. A
heavy quark expansion in terms of the inverse of the dressed heavy
quark mass $1/\hat{m}_Q$ is presented with a complete decomposition
of the full field and integrating out the small components. The
heavy quark-antiquark coupling effects are included in the finite
mass corrections. It is shown that the $1/\hat{m}_Q$ expansion is
more favorable in application. The extraction of $|V_{cb}|$ from
exclusive B decays is studied by using such a new expansion
approach.

\end{abstract}

\maketitle

\section{Introduction}

For a heavy hadron which contains a single heavy quark (bottom or
charm), the heavy quark mass $m_Q$ is much larger than the QCD
energy scale $\Lambda_{QCD}$ which characterizes the light
degrees of freedom in the heavy hadron. The four momentum of the heavy quark
can be expressed as $p^\mu =m_Q v^\mu +k^\mu $, where $v^\mu$ is
taken to be the velocity of hadron at the rest frame, and $k^\mu$ is
the residual momentum of the order of binding energy, which is much
smaller than $m_Q$. The heavy quark symmetry
\cite{symmetry,symmetry2,Is.Wi} and its breaking effects are of
particular importance in studying such hadrons. Consequently, the
heavy quark effective theory (HQET) has been developed, where the
effective Lagrangian is expanded in $1/m_Q$. In deriving the Lagrangian of HQET,
the quark and antiquark are assumed to be conserved separately. Namely the
heavy antiquarks are regarded as completely decoupled from the heavy
quarks at the beginning. The transition matrix elements can also be
represented in series of $1/m_Q$ through the heavy quark expansion
(HQE) and evaluated order by order. HQET and HQE have been discussed
by many authors
\cite{MBVoloshin292,HGeorgi447,TMannel2388,MELuke447,BGrinstein253,AFFalk1,AFFalk185,TMannel204,BGrinstein34,TMannel428,TMannel396,MNeubert259}.
In the past two decades they are widely used in
studying heavy hadrons.

When the momentum of the heavy quark is much lower than the quark-antiquark
pair creation threshold, an alternative framework of effective field
theory for heavy quarks can directly be derived from the full QCD
\cite{YLWu819,WYWang1817,YLWu1303,YLWu5743}. Just like for other
effective theories, the basic idea is that some degrees of freedom
characterizing higher scale physics can be decomposed and integrated
out when we consider physics at low energy scales. Concretely
speaking, for heavy quarks with $|{\bf p}|\ll 2m_Q$,  one may
perform a complete decomposition of the QCD full field into quark
field and antiquark field via positive and negative energy
components (see below) of a full field, and integrate out the small
components of quark field and antiquark field, which leads to the
so-called $1/m_Q$ corrections. When considering heavy quark (or
antiquark) systems, one should further integrate in the
contributions of heavy antiquark (or quark) components. As a
consequence, additional $1/m_Q$ corrections arise from the
quark-antiquark coupling terms in the full QCD. It should be noted
that such a framework is distinguishable from the usual HQET in
which the particle and antiparticle were assumed to be conserved
separately and treated in a different way. For convenience, we refer
to such a framework as a heavy quark effective field theory (HQEFT).
Though the heavy quark-antiquark coupling effects vanish in the
heavy quark limit, they are actually nonzero when one considers the
finite mass contributions. It is then not surprising that the
$1/m_Q$ corrections evaluated in HQEFT and HQET could be different
though they are the same in the infinity mass limit. For instance,
the transition matrix elements of $1/m_Q$ corrections concern less
independent wave functions in HQEFT than in the usual HQET and the
$1/m_Q$ order corrections at zero recoil automatically vanish in HQEFT,
and there exist some relations between wave functions and heavy
hadron masses in HQEFT.

HQEFT has been applied to explore various processes of heavy
hadrons. In particular, the Cabibbo-Kobayashi-Maskawa (CKM) matrix
elements $|V_{cb}|$ and $|V_{ub}|$ are extracted from both inclusive
\cite{YAYan2735,YLWu285,YBZuo3685,WYWang1379} and exclusive
\cite{WYWang1817,WYWang377,WYWang57,WYWang219,WYWang2743,WYWang014024,WYWang228}
B decays. In the treatment of inclusive B decays, the dressed heavy
quark mass $\hat{m}_Q=m_Q+\bar{\Lambda}$ as a whole enters the
formulation, which implies that a ``dressed heavy quark"-hadron
duality is more reasonable than the naive heavy quark-hadron
duality. As a consequence, when the inclusive decay rates of heavy
hadrons are expressed in terms of the physical hadron masses, they
receive no $1/\hat{m}_Q$ order corrections. This treatment not only
suppresses the next-to-leading order contributions and makes the
operator product expansion (OPE) reliable, but also diminishes the
large uncertainties arising from the heavy quark mass. For exclusive
decays, HQEFT has also been demonstrated to be reliable. Whereas in
our previous works the heavy quark expansion for the effective
Lagrangian and transition matrix elements is carried out in powers
of $1/m_Q$.

In this paper, we briefly review the description of HQEFT and show
that a $1/\hat{m}_Q$ expansion is also consistently applicable to
the heavy quark effective Lagrangian and transition matrix elements.
In Sec. \ref{SectionLagrangian}, we first outline the derivation of
a complete HQEFT and then extend it to the formulation in terms of
$1/\hat{m}_Q$ expansion. In Sec. \ref{Sectionmatixelement}, we
present new formulae for HQE of heavy-to-heavy transition matrix
elements by applying the $1/\hat{m}_Q$ expansion. In Sec.
\ref{Sectionvcb}, we extract the CKM matrix element $|V_{cb}|$ based
on the new formulation and the most recent experimental data. Our
conclusions are given in Sec. \ref{Sectionsummary}.

\section{Heavy Quark Expansion in Terms of $1/\hat{m}_Q$}\label{SectionLagrangian}

The Lagrangian in the full QCD is
\begin{eqnarray}
{\cal L}_{QCD}=\bar{Q}(i\DS -m_Q)Q+{\cal L}_{light} .
\end{eqnarray}
$Q$ is the full field for heavy quark, and ${\cal L}_{light}$
represents the section containing no heavy quarks. Based on the
principle of superposition, the field $Q$
in quantum field theory is actually the superposition of two parts
which correspond to the positive and negative energy components.
Namely the full field $Q$ can always be decomposed formally into
positive and negative energy parts
\begin{eqnarray}
\label{Qzf} Q=Q^{+} + Q^{-} ,
\end{eqnarray}
where $Q^+$ and $Q^-$ may be expressed explicitly in the energy-momentum space as
\begin{eqnarray}
\label{Qplus} Q^{+}(x)
&  =&  \int \frac{d^4p}{(2\pi)^4} \theta(p^0) Q^{+}(p)e^{-ip\cdot x}, \\
\label{Qminus}
 Q^{-}(x)
  &=& \int \frac{d^4p}{(2\pi)^4}   \theta(p^0) Q^{-}(p) e^{ip\cdot x}
   =  \int \frac{d^4p}{(2\pi)^4} \theta(-p^0) Q^{-}(-p) e^{-ip\cdot x} .
\end{eqnarray}
It is clear that $Q^+$ and $Q^-$ correspond to the positive and
negative energy parts of the full field $Q$, which are the so-called
quark field and antiquark field, respectively. In the case for
free quark fields, they can be expanded in terms of plane
waves as
\begin{eqnarray}
\label{Qzexp}
Q^{+}(x) &=& \int \frac{d^3p}{(2\pi)^3}{m\over E}
    \sum\limits_s b_s(p)u_s(p)e^{-ip\cdot x} \nonumber \\
    & = & \int \frac{d^4p}{(2\pi)^4} (2\pi) 2m \delta(p^2-m^2) \theta(p^0) \sum\limits_s b_s(p)u_s(p)e^{-ip\cdot x} ,  \\
\label{Qfexp}
Q^{-}(x) &=& \int \frac{d^3p}{(2\pi)^3}{m\over E}
    \sum\limits_s d^{\dag}_s(p)v_s(p)e^{ip\cdot x} \nonumber \\
    & = & \int \frac{d^4p}{(2\pi)^4} (2\pi) 2m \delta(p^2-m^2) \theta(p^0)  \sum\limits_s d^{\dag}_s(p)v_s(p)e^{ip\cdot
    x} ,
\end{eqnarray}
where $s$ is the spin index, $b_s$ and $d^\dagger_s$ are the
annihilation and creation operators respectively, and $u_s$ and
$v_s$ are four-component spinors. In Dirac representation, they can
be explicitly written as
\begin{eqnarray}
u_s(p) &=& \sqrt{p^0 + m \over 2m} \left( \begin{array}{c} 1 \\
{{\bf \sigma \cdot p}\over p^0+m} \end{array} \right)
\varphi_s ,  \\
v_s(p) &=& \sqrt{p^0 + m \over 2m} \left( \begin{array}{c} {{\bf
\sigma \cdot p}\over p^0+m} \\ 1 \end{array} \right) \chi_s
\end{eqnarray}
with $\varphi_s$ being the two component Pauli spinor field that
annihilates a heavy quark, and $\chi_s$ being the Pauli spinor field
that creates a heavy antiquark. Later on we will see that
the effective heavy quark and antiquark fields at
$|{\bf p}| \ll 2 m_Q$ exactly come from the ``large" components of
$Q^+$ and $Q^-$.

The generating functional in the full theory can be represented as
\begin{eqnarray}
\label{generatingfun1}
Z[j]&=&\int {\cal D}q {\cal D}Q e^{i\int d^4x ({\cal
L}_{QCD}[q,Q]+j \phi )} .
\end{eqnarray}
In Eq.(\ref{generatingfun1}) and all the following relevant
equations, we simply denote the source terms as $j\phi$ for
convenience.

Introducing a vector $v^\mu$ with $v^2 = 1$, one can define the
projecting operators
\[ P_{\pm} = \frac{1\pm \VS}{2} , \]
which satisfies
\[  P_{\pm}^2 = P_{\pm} . \]
Then $Q^{\pm}$ can be written as
\begin{eqnarray}
\label{Qdec1} Q^{+} & = & \Big({1 + \VS \over 2} + {1 -
\VS \over 2} \Big) Q^{+}
= \hat{Q}^{+}_v + R^{+}_v,  \\
\label{Qdec2} Q^{-} & = & \Big({1 - \VS \over 2} + {1 +
\VS \over 2}\Big) Q^{-} = \hat{Q}^{-}_v + R^{-}_v
\end{eqnarray}
with
    \begin{equation}
    \label{eq:9}
      \hat{Q}^{\pm}_v\equiv \frac{1 \pm \VS }{2}Q^{\pm},\hspace{1.5cm}
      R^{\pm}_v\equiv \frac{1 \mp \VS }{2}Q^{\pm} .
    \end{equation}

At $|{\bf p}| \ll 2 m_Q$, the field components $R^{+}_v$ and $R^{-}_v$ become
``small components" of quarks and antiquarks, while $\hat{Q}_v^{+}$
and $\hat{Q}_v^{-}$ are the ``large components" \cite{YLWu1303,YLWu5743}.
To be more explicit, taking $v=(1, 0, 0,0)$, one then has in the momentum space
\begin{eqnarray}
\label{component1} \hat{Q}_v^{+} \to \PP u_s(p) &=& \sqrt{p^0 + m
\over 2m} \left( \begin{array}{c} 1 \\ 0 \end{array} \right)
\varphi_s ,
\\
\label{component2} R_v^{+} \to \PM u_s(p) &=& \sqrt{p^0 + m \over
2m} \left( \begin{array}{c} 0 \\ {{\bf \sigma \cdot p}\over p^0 +
m}
\end{array} \right) \varphi_s ,
\\
\label{component3} R_v^{-} \to \PP v_s(p) &=& \sqrt{p^0 + m \over
2m} \left( \begin{array}{c}{{\bf \sigma \cdot p}\over p^0 + m} \\
0
\end{array} \right) \chi_s ,
\\
\label{component4} \hat{Q}^{-}_v \to \PM v_s(p) &=& \sqrt{p^0 + m
\over 2m} \left( \begin{array}{c} 0 \\ 1 \end{array} \right) \chi_s
.
\end{eqnarray}
In this case one can decompose the full field $Q$ as
Eqs.(\ref{Qzf}), (\ref{Qdec1}) and (\ref{Qdec2}), and write the generating functional as
\begin{eqnarray}
\label{generatingfun2} Z[j] =
   \int {\cal D}q {\cal D}\hat{Q}^{+}_v {\cal D}\hat{Q}^{-}_v {\cal D}R^{+}_v {\cal D}R^{-}_v  e^{i \int d^4x ({\cal
L}_{QCD}[q,\hat{Q}^{+}_v ,\hat{Q}^{-}_v, R^{+}_v, R^{-}_v]
+j \phi )}  .
\end{eqnarray}

Then one may integrate out the small components $R^{+}_v$
and $R^{-}_v$ to get
\begin{eqnarray}
\label{generatingfun3}
 Z[j]=
  \int {\cal D}q {\cal D}\hat{Q}^{+}_v {\cal D}\hat{Q}^{-}_v   e^{i\int d^4x ({\cal L}_{light}+\hat{\cal
L}_{Q,v}[\hat{Q}^{+}_v ,\hat{Q}^{-}_v ] +j \phi )}    .
\end{eqnarray}
$\hat{\cal L}_{Q,v}$ is the resulting Lagrangian for heavy section
with the small components integrated out. It can also be derived
equivalently by using the relevant Dirac equation of motion
\begin{eqnarray}
(i\DSP - m_Q) R_v^{\pm}  + i\DSC \hat{Q}_v^{\pm} = 0 , \\
\bar{R}^{\pm}_v (-i \overleftarrow \DSP - m_Q)  - \QVHPMB i
\overleftarrow \DSC = 0 ,
\end{eqnarray}
where $\DSP $, $\DSC $, $\overleftarrow \DSP $, $\overleftarrow \DSC $ and
$\overleftarrow D^\mu $ are defined as
\begin{eqnarray}
 \label{eq:12}
&&  \DSP = \VS v\cdot D, \hspace{2cm}
  \DSC = \DS-\VS v\cdot D,  \hspace{2cm}
  \overleftarrow \DSP = \VS v\cdot \overleftarrow D , \nonumber \\
&&  \overleftarrow \DSC = \overleftarrow \DS -
     \VS v \cdot \overleftarrow D   , \hspace{1cm}
 \int\kappa \overleftarrow D^\mu \varphi
   = -\int\kappa D^{\mu}\varphi .
\end{eqnarray}
Clearly one has
\begin{eqnarray}
\label{relation} \{ \vslash , \DSC \} = [\vslash, \DSP ] = 0.
\end{eqnarray}

$\hat{\cal L}_{Q,v}$ is found to be \cite{WYWang1817,YLWu5743}
    \begin{equation}
     \label{LhatQv}
        \hat{{\cal L}}_{Q,v} = \hat{{\cal L}}^{(++)}_{Q,v}+\hat{\cal L}^{(--)}_{Q,v}+\hat{\cal L}^{(+-)}_{Q,v}+\hat{\cal L}^{(-+)}_{Q,v}
     \end{equation}
  with
    \begin{eqnarray}
    \label{Lzfzf}
      \hat{\cal L}^{(\pm \pm)}_{Q,v} &=& \QVHPMB [i \hat {\cal \not \!\! D}_{\! v} -\MQ
           ] \QVHPM  ,   \nonumber \\
      \hat{\cal L}^{(\pm \mp)}_{Q,v} &=& \frac{1}{2m_Q} \QVHPMB
       ( -i \overleftarrow{ \hat {\cal \not \!\! D}}_{\! v}
        -m_Q )  \Big( 1-\frac{i\DSP
         +\MQ}{2\MQ} \Big)^{-1} (i\DSC)  \QVHMP \nonumber \\
         & = & \frac{1}{2m_Q}
    \QVHPMB ( -i  \overleftarrow{\DS}_{\! \bot})
           \Big( 1-\frac{-i \overleftarrow \DSP
         +\MQ}{2\MQ} \Big)^{-1}
           ( i \hat {\cal \not \!\! D}_{\! v}-m_Q)  \QVHMP ,
    \end{eqnarray}
where $i\hat{\cal \not \! \!D}_{\! v} $ is defined as
\begin{eqnarray}
i\hat{\cal \not \! \!D}_{\! v} = i\DSP +\frac{1}{2m_Q} i\DSC
 \Big( 1-\frac{i\DSP    +\MQ}{2\MQ} \Big)^{-1}  i\DSC  ,
\end{eqnarray}
and the operator $-i\overleftarrow{ \hat{\cal \not \!\! D}}_{\! v}$ can
be obtained from $ i\hat{\cal \not \!\! D}_{\! v}$ by replacing
$D^\mu$ with $-\overleftarrow{D}^\mu$.

To get a reliable expansion at low energies, the large momentum carried by the heavy
quark should be removed. Generally this can be achieved by
introducing new field variables
    \begin{equation}
    \label{momentumshift}
       Q_v=e^{iv\hspace{-0.15cm}\slash \hat{m}_Q v\cdot x}\QVH ,\hspace{1.5cm}
       \bar{Q}_v=\QVHB e^{-i v\hspace{-0.15cm}\slash \hat{m}_Q v \cdot x} .
    \end{equation}
$\hat{m}_Q$ is a parameter with mass dimension. It can be chosen
appropriately according to the physical picture for the process
studied. In general a heavy quark within a hadron cannot truly be on
shell due to strong interaction among heavy and light quarks as
well as soft gluons. Thus one may write the total momentum of the
heavy quark in a hadron as $P_Q=m_Q v + k = \hat{m}_Q v + \tilde{k}
$, where $v\cdot\tilde{k}$ is the part which depends on heavy flavor
and is suppressed by the heavy quark mass; $\hat{m}_Q $ is defined
as the sum of the heavy quark mass and the binding energy
$\bar{\Lambda}$ that reflects the nonperturbative effects of strong
interaction and relates to the light constituents in the heavy
hadron, $\hat{m}_Q=m_Q+\bar{\Lambda}$. In such a consideration, the residual
momentum $k=\bar{\Lambda}v + \tilde{k}$ as a whole characterizes the
off-shellness of the heavy quark in the heavy hadron. Namely, the
total residual momentum $k = \bar{\Lambda}v +\tilde{k} $ of the
heavy quark is assumed to comprise the main contributions of the
light degrees of freedom in the heavy hadron containing a single
heavy quark. In such a physical picture, the heavy quark may be
regarded as a ``dressed heavy quark", and the heavy hadron
containing a single heavy quark is more reliable to be considered as
a dualized particle of a ``dressed heavy quark". Different from the
``heavy quark"-hadron duality in the usual heavy quark effective
theory, what we are considering is the physical picture of the
``dressed heavy quark"-hadron duality. As a consequence, the wave
functions defined in the next section should have a weaker
dependence on the light constituents of heavy hadrons.

With the definition (\ref{momentumshift}), the Lagrangian
(\ref{LhatQv}) can be written in terms of $\QV$ and $\QVB$, which
carry only the small residual momentum $\tilde{k}^\mu=p^\mu_H-\hat{m}_Q v^\mu$. Explicitly, one has\cite{YLWu5743}
 \newcommand{\CDSv}{{\cal D}\hspace{-0.25cm}\slash_{v}}
\begin{eqnarray}
 \label{LQv}
 {\cal L}_{Q,v}&=& {\cal L}^{I}_{Q,v}+{\cal L}^{II}_{Q,v}
   = {\cal L}^{(0)}_{Q,v}+{\cal L}^{(1/\hat{m}_Q)}_{Q,v}  ,  \\
 \label{LQv1}
 {\cal L}^{I}_{Q,v}&=&  {\cal L}^{(++)}_{Q,v}+ {\cal L}^{(--)}_{Q,v} = \bar{Q}_v(i\CDSv ) Q_v
 \equiv {\cal L}^{(0)}_{Q,v}+{\cal L}^{I(1/\hat{m}_Q)}_{Q,v},  \\
 \label{LQv2}
 {\cal L}^{II}_{Q,v} &=& {\cal L}^{(+-)}_{Q,v}+ {\cal L}^{(-+)}_{Q,v} \nonumber \\
    &=& \frac{1}{2\hat{m}_Q} \bar{Q}_v (-i\overleftarrow \CDSv )
    e^{2i v\hspace{-0.15cm}\slash \hat{m}_Q v\cdot x}\Big( 1-\frac{i\DSP+\bar{\Lambda}}{2\hat{m}_Q}\Big)^{-1}
    (i\DSC) Q_v   \nonumber \\
 &=& \frac{1}{2\hat{m}_Q} \bar{Q}_v (-i\overleftarrow \DSC)
   \Big(1-\frac{-i\overleftarrow \DSP+\bar{\Lambda}}{2\hat{m}_Q}\Big)^{-1}
   e^{-2i v\hspace{-0.15cm}\slash \hat{m}_Q v\cdot x} (i\CDSv ) Q_v
  \nonumber  \\
   & \equiv & {\cal L}^{II(1/\hat{m}_Q)}_{Q,v}
\end{eqnarray}
 with
\begin{eqnarray}
 i {\cal \not \! \!D}_{\! v} &=& i\DSP+\bar{\Lambda}
   +\frac{1}{2\hat{m}_Q} i\DSC
 \Big( 1-\frac{i\DSP+\bar{\Lambda}}{2\hat{m}_Q} \Big)^{-1}  i\DSC ,
   \nonumber \\
 -i \overleftarrow{\cal \not \!\! D}_{\! v} &=& -i\overleftarrow \DSP
  +\bar{\Lambda}+\frac{1}{2\hat{m}_Q} (-i\overleftarrow \DSC)
 \Big( 1-\frac{-i\overleftarrow \DSP+\bar{\Lambda}}{2\hat{m}_Q} \Big)^{-1}  (-i\overleftarrow\DSC) .
\end{eqnarray}
In Eq.(\ref{LQv}) we use ${\cal L}^{(0)}_{Q,v}$ to denote the leading term in the
$1/\hat{m}_Q$ expansion of the Lagrangian ${\cal L}_{Q,v}$, and
${\cal L}^{(1/\hat{m}_Q)}_{Q,v} $ contains all $1/\hat{m}_Q$ corrections
to ${\cal L}^{(0)}_{Q,v}$. From Eqs.(\ref{LQv})-(\ref{LQv2}) one has
\begin{eqnarray}
 \label{LQv0}
 {\cal L}^{(0)}_{Q,v}&=& \bar{Q}_v (i\DSP+\bar{\Lambda}) Q_v ,\\
 {\cal L}^{I(1/\hat{m}_Q)}_{Q,v}&=&\bar{Q}_v
   \frac{1}{2\hat{m}_Q} i\DSC
 \Big( 1-\frac{i\DSP+\bar{\Lambda}}{2\hat{m}_Q} \Big)^{-1}  i\DSC
 Q_v, \\
{\cal L}^{(1/\hat{m}_Q)}_{Q,v}&=&{\cal L}^{I(1/\hat{m}_Q)}_{Q,v}
    +{\cal L}^{II(1/\hat{m}_Q)}_{Q,v} .
\end{eqnarray}
Note that
\begin{eqnarray}
Q_v=Q^+_v + Q^-_v ,
\end{eqnarray}
so the effective
Lagrangian ${\cal L}_{Q,v}$ is complete for the large component of heavy quark and antiquark. In deriving ${\cal L}_{Q,v}$ we have only integrated over the small component ($R_v=R^+_v+R^-_v$) of heavy quark and antiquark fields.

In the above discussions, the binding energy $\bar{\Lambda}$ is
introduced based on physical consideration of the heavy-light
systems. Formally it can be defined consistently via the
normalization of hadron states as follows. The hadron state
$|H\rangle$ in full theory is normalized as
      \begin{equation}
      \label{normalizationQCD}
         \langle H(p)| \bar{Q}\gamma^{\mu}Q | H(p)\rangle =2p_H^{\mu}=2m_H v^{\mu},
      \end{equation}
where $p_H^{\mu}=m_H v^{\mu}$ is the momentum of the heavy hadron $H$.
In the effective theory at low energies, one may introduce an
effective heavy hadron state $| H_v \rangle $, which is heavy flavor
independent, and normalized as
\begin{equation}
     \label{normalizationHQEFT}
   \langle H_{v} | \bar{Q}_v \gamma^{\mu} Q_v | H_v \rangle = 2\bar{\Lambda} v^{\mu} .
\end{equation}
It is then related to the heavy hadron state $| H\rangle $ via
    \begin{equation}
    \label{matrixHHv}
     \frac{1}{\sqrt{m_{H^{\prime}}m_{H}}} \langle H^{\prime} | \bar{Q}^{\prime} \Gamma Q | H\rangle=
\frac{1}{\sqrt{\bar{\Lambda}_{H^{\prime}} \bar{\Lambda}_H}}
\langle H^{\prime}_{v^{\prime}}|
          J_{Q,v} e^{i\int d^4x {\cal L}^{(1/\hat{m}_Q)}_{Q,v}}| H_v\rangle ,
    \end{equation}
where $\Gamma$ denotes Dirac matrixes, $  \bar{\Lambda}_{H} = m_{H}-m_{Q}$ and $ \bar{\Lambda}_{H'} =
m_{H'}-m_{Q'}  $ are the mass differences between heavy hadrons and
heavy quarks, while
\[ \bar{\Lambda} = \lim_{m_{Q}\to \infty} \bar{\Lambda}_H   \]
is independent of the heavy flavor and reflects the contributions of
light degrees of freedom in the hadron.
$J_{Q,v}$ in Eq.(\ref{matrixHHv}) is derived from the current
 $\bar{Q}\Gamma Q$. It will be given explicitly in Sec.\ref{Sectionmatixelement}.

The basic framework of HQEFT has been derived and discussed in the previous
papers\cite{YLWu819,WYWang1817,YLWu1303,YLWu5743}. Here we reexpress
the effective Lagrangian
in terms of the expansion $1/\hat{m}_Q$ instead of $1/m_Q$. The
binding energy arising from the ``longitudinal residual momentum" of
the heavy quark is absorbed into the heavy quark mass to be given as
the dressed heavy quark mass $\hat{m}_Q= m_Q + \bar{\Lambda}$, so
that the flavor independent nonperturbative contributions of the
light degrees of freedom are effectively included in the dressed
heavy quark. This treatment is consistent with the physical picture
of a heavy kernel (the dressed heavy quark) surrounded with the
clouds of light degrees of freedom which mainly reflect the small
``transverse residual momentum" of heavy quark. As $\hat{m}_Q >
m_Q$, an expansion in $1/\hat{m}_Q$ is expected to be more
convergent and reliable, especially for charm quark systems.
Actually, it has been shown in the inclusive
decays\cite{YAYan2735,YLWu285} that by using $\hat{m}_Q$ instead of
adopting $m_Q$ and $\bar{\Lambda}$ separately in the calculations,
the results get less uncertainties and the order $1/\hat{m}_Q$
corrections are automatically absent. In this note, we will show
that based on the new formulation of Lagrangian given in Eqs.(\ref{LQv})-(\ref{LQv2}), one can also consistently perform
$1/\hat{m}_Q$ expansion for the exclusive decays.% of the heavy

\section{ Transition Matrix Elements in $1/\hat{m}_Q$ expansion}\label{Sectionmatixelement}

Similar to the derivation of the effective Lagrangian ${\cal L}_{Q,v}$, the heavy quark current $J(x)=\bar{Q}'(x)\Gamma Q(x)$ in full QCD can also be transformed into the following form by integrating out the small component $R_v$,
\begin{eqnarray}
J(x) \to J_{Q,v}(x) &=& \bar{\hat{Q}}'_{v'}(x) \Gamma \hat{Q}_v(x)
+\bar{\hat{Q}}'_{v'}(x) \Gamma \hat{W}_v \hat{Q}_v(x)  \nonumber \\
&&+ \bar{\hat{Q}}'_{v'}(x) \overleftarrow{\hat{W}}_{v'} \Gamma \hat{Q}_v(x)
+\bar{\hat{Q}}'_{v'}(x) \overleftarrow{\hat{W}}_{v'} \Gamma \hat{W}_v
   \hat{Q}_v(x) \nonumber \\
 &=& \bar{Q}'_{v'}(x) e^{iv\hspace{-0.15cm}\slash ' \hat{m}_{Q'} v'\cdot x}
   \Gamma    e^{-iv\hspace{-0.15cm}\slash \hat{m}_Q v\cdot x}   Q_v(x) \nonumber\\
&&+\bar{Q}'_{v'}(x) e^{iv\hspace{-0.15cm}\slash ' \hat{m}_{Q'} v'\cdot x}
   \Gamma   e^{iv\hspace{-0.15cm}\slash \hat{m}_Q v\cdot x}
   W_v Q_v(x) \nonumber \\
&&+ \bar{Q}'_{v'}(x) \overleftarrow W_{v'}
   e^{-iv\hspace{-0.15cm}\slash ' \hat{m}_{Q'} v'\cdot x}
   \Gamma e^{-iv\hspace{-0.15cm}\slash \hat{m}_Q v\cdot x}  Q_v(x) \nonumber \\
&&+ \bar{Q}'_{v'}(x) \overleftarrow W_{v'}
   e^{-iv\hspace{-0.15cm}\slash ' \hat{m}_{Q'} v'\cdot x}
   \Gamma e^{iv\hspace{-0.15cm}\slash \hat{m}_Q v\cdot x}
   W_v  Q_v(x) ,
\end{eqnarray}
where
\begin{eqnarray}
\hat{W}_v &=& \frac{1}{2m_Q} \Big(1-\frac{i\DSP+m_Q}{2m_Q}\Big)^{-1} i\DSC, \nonumber\\
W_v &=& \frac{1}{2\hat{m}_Q}\Big(1-\frac{i\DSP+\bar{\Lambda}}{2\hat{m}_Q}\Big)^{-1} i\DSC ,\nonumber\\
\overleftarrow {\hat{W}_{v'}} &=& \frac{1}{2m_{Q'}} (-i\overleftarrow\DSC)
 \Big(1-\frac{-i\overleftarrow\DSP+m_{Q'}}{2m_{Q'}}\Big)^{-1} , \nonumber\\
\overleftarrow W_{v'} &=& \frac{1}{2\hat{m}_{Q'}} (-i\overleftarrow\DSC)
 \Big(1-\frac{-i\overleftarrow\DSP+\bar{\Lambda}'}{2\hat{m}_{Q'}}\Big)^{-1}
\end{eqnarray}
with
\begin{eqnarray}  \overleftarrow \DSP = \VS' v'\cdot \overleftarrow D , \hspace{2cm}
  \overleftarrow \DSC = \overleftarrow \DS -
     \VS' v' \cdot \overleftarrow D   ,
\end{eqnarray}
which have minor differences to the formulae in (\ref{eq:12}).

Using Eq.(\ref{relation}), and noting that
\begin{eqnarray}
\label{relation1} e^{iA v\hspace{-0.15cm}\slash }\frac{1+\sigma
\vslash}{2} &=& e^{iA\sigma }\frac{1+\sigma \vslash}{2},
\end{eqnarray}
where $\sigma =\pm 1$ and $A$ is a c-number, we write $J_{Q,v}$ as
\begin{eqnarray}
J_{Q,v}(x)
&=& \sum _{\sigma, \sigma ^{\prime} = \pm 1}e^{i (\sigma ^{\prime}
\hat{m}_{Q'}v' - \sigma \hat{m}_{Q}v)\cdot x}
\bar{Q}'_{v'}(x)\frac{1+\sigma ^{\prime} \vslash ' }{2} (\Gamma
 + \Gamma W_v \nonumber \\
 &&+ \overleftarrow W_{v'} \Gamma +\overleftarrow W_{v'}
\Gamma W_v )\frac{1+\sigma \vslash }{2}Q_v(x).
\end{eqnarray}
Furthermore, both $J_{Q,v}$ and ${\cal L}_{Q,v}$ can be expanded in terms of
$1/\hat{m}_{Q^{(')}}$. Explicitly one obtains
\begin{eqnarray}
\label{JQvexp}
J_{Q,v}(x) &=& J_{Q,v}^{(0)}(x)+J_{Q,v}^{(1/\hat{m}_Q)}(x), \\
\label{JQvexp0}
J_{Q,v}^{(0)}(x) & = & \sum _{\sigma, \sigma ^{\prime} = \pm 1}e^{i
(\sigma ^{\prime} \hat{m}_{Q'}v' - \sigma \hat{m}_{Q}v)\cdot x}
\bar{Q}'_{v'}(x)\frac{1+\sigma ^{\prime} \vslash ' }{2}\Gamma
\frac{1+\sigma \vslash }{2}Q_v(x), \\
\label{JQvexpcor}
 J_{Q,v}^{(1/\hat{m}_Q)}(x)& = &
\sum _{\sigma, \sigma ^{\prime} = \pm 1}e^{i (\sigma ^{\prime}
\hat{m}_{Q'}v' - \sigma \hat{m}_{Q}v)\cdot x}
\bar{Q}'_{v'}(x)\frac{1+\sigma ^{\prime} \vslash ' }{2}
\Big [ \frac{1}{2\hat{m}_Q} \Gamma i\DSC
+\frac{1}{2\hat{m}_{Q'}}(-i\overleftarrow\DSC)\Gamma \nonumber \\
&&+\frac{1}{4\hat{m}_{Q}^{2}}\Gamma(i\DSP+\bar{\Lambda})i\DSC
+\frac{1}{4\hat{m}_{Q'}^{2}}(-i\overleftarrow \DSC)(-i\overleftarrow\DSP+\bar{\Lambda}')\Gamma \nonumber\\
&&+ \frac{1}{4\hat{m}_{Q}\hat{m}_{Q'}}(-i\overleftarrow \DSC)\Gamma
  (i\DSC)
  + O\Big ( \frac{1}{\hat{m}_{Q^{(\prime )}}^{3}}\Big)
    \Big ]
\frac{1+\sigma \vslash }{2}Q_v(x)
\end{eqnarray}
and
\begin{eqnarray}
\label{LQvexp1}
{\cal L}^{I(1/\hat{m}_Q)}_{Q,v} & = &
 \sum _{\varepsilon = \pm 1} \bar{Q}_v \frac{1+ \varepsilon \vslash }{2}
\Big[\frac{(i\DSC)^{2}}{2\hat{m}_Q}+ \frac{1}{4\hat{m}_{Q}^{2}}i\DSC
(i\DSP +\bar{\Lambda})i\DSC  \nonumber\\
&&+ O\Big(
\frac{1}{\hat{m}_Q^3} \Big) \Big]
\frac{1+ \varepsilon \vslash }{2}  Q_v  , \\
\label{LQvexp2}
{\cal L}^{II(1/\hat{m}_Q)}_{Q,v} & = & \sum _{\varepsilon = \pm 1}
e^{2i\varepsilon \hat{m}_{Q}v\cdot x }\bar{Q}_v \frac{1+ \varepsilon
\vslash }{2}\Big[ \frac{1}{2\hat{m}_Q}(-i \overleftarrow\DSP +\bar{\Lambda
})i\DSC \nonumber \\
&& +\frac{1}{4\hat{m}_{Q}^{2}}(-i \overleftarrow\DSP +\bar{\Lambda })
(i\DSP +\bar{\Lambda})i\DSC \nonumber \\
&&+\frac{1}{4\hat{m}_{Q}^{2}}(-i\overleftarrow \DSC)^{2}i\DSC
  + O\Big(\frac{1}{\hat{m}_Q^3} \Big)
\Big]\frac{1- \varepsilon \vslash }{2} Q_v  .
\end{eqnarray}

Then the effective current in terms of $1/\hat{m}_Q$ expansion is obtained:
\begin{eqnarray}
 \label{neweffectivecurrent}
  J^{eff}_{Q,v}(x) &\equiv & \langle J_{Q,v}(x) e^{i\int d^4y {\cal L}^{(1/\hat{m}_Q)}_{Q,v}} \rangle \nonumber\\
  &=& \sum _{\sigma, \sigma ^{\prime} = \pm 1} e^{i (\sigma ^{\prime} \hat{m}_{Q'}v' - \sigma \hat{m}_{Q}v)\cdot x} \bar{Q'}_{v'} \frac{1+\sigma ^{\prime} \vslash ' }{2} \Big [ \Gamma - \frac{1}{2\hat{m}_Q}O_{1}(\Gamma) - \frac{1}{2\hat{m}_{Q'}}O_{1}^{\prime}(\Gamma) \nonumber \\
  && -\frac{1}{4\hat{m}_Q^2}O_{2}(\Gamma) -\frac{1}{4\hat{m}_{Q'}^2}O_{2}^{\prime}(\Gamma)   + \frac{1}{4\hat{m}_Q^2}O_{3}(\Gamma) + \frac{1}{4\hat{m}_{Q'}^2}O_{3}^{\prime}(\Gamma) \nonumber\\
  &&  +\frac{1}{4\hat{m}_{Q'} \hat{m}_Q} O_{4}(\Gamma) + O\Big( \frac{1}{\hat{m}_{Q^{(\prime)}}^3} \Big) \Big ]\frac{1+\sigma \vslash }{2}Q_{v} ,
  \end{eqnarray}
where the operators are defined as
 \begin{eqnarray}
   \label{eq:44}
     O_1(\Gamma)& =&\Gamma\frac{1}{i\DSP+\bar{\Lambda}}(i\DSC)^2 ,\nonumber\\
     O_1^{\prime}(\Gamma)&=&(-i\stackrel{\hspace{-0.1cm}\leftarrow}{\DSC})^2
\frac{1}{-i\stackrel{\hspace{-0.1cm}\leftarrow}
        {\DSP}+\bar{\Lambda}'} \Gamma ,   \nonumber \\
     O_2(\Gamma)& =&\Gamma\frac{1}{i\DSP+\bar{\Lambda}}(i\DSC)(i\DSP+\bar{\Lambda})i\DSC , \nonumber\\
     O_2^{\prime}(\Gamma)& =&(-i\stackrel{\hspace{-0.1cm}\leftarrow}{\DSC})
(-i\stackrel{\hspace{-0.1cm}\leftarrow}{\DSP} +\bar{\Lambda}')
        (-i\stackrel{\hspace{-0.1cm}\leftarrow}{\DSC})
\frac{1}{-i\stackrel{\hspace{-0.1cm}\leftarrow}{\DSP}+\bar{\Lambda}'}
        \Gamma ,\nonumber\\
     O_3(\Gamma) &=&\Gamma\frac{1}{i\DSP+\bar{\Lambda}}(i\DSC)^2 \frac{1}{i\DSP+\bar{\Lambda}}
        (i\DSC)^2 , \nonumber\\
     O_3^{\prime}(\Gamma)& =&(-i\stackrel{\hspace{-0.1cm}\leftarrow}{\DSC})^2
\frac{1}{-i\stackrel{\hspace{-0.1cm}\leftarrow}
        {\DSP}+\bar{\Lambda}'}(-i\stackrel{\hspace{-0.1cm}\leftarrow}{\DSC})^2
\frac{1}{-i\stackrel{\hspace{-0.1cm}\leftarrow}
        {\DSP}+\bar{\Lambda}'} \Gamma ,\nonumber\\
     O_4(\Gamma)& =&(-i\stackrel{\hspace{-0.1cm}\leftarrow}{\DSC})^2
\frac{1}{-i\stackrel{\hspace{-0.1cm}\leftarrow}
        {\DSP}+\bar{\Lambda}'}\Gamma\frac{1}{i\DSP+\bar{\Lambda}}(i\DSC)^2.
   \end{eqnarray}
Eq.(\ref{neweffectivecurrent}) can be derived by using
\begin{equation}
\label{propagator}
  \frac{i}{i\DSP+\bar{\Lambda}}
\end{equation}
as the propagator when $Q_v$ and $\bar{Q}_v$ fields are contracted. The feasibility of this treatment is shown in the appendix.

Note that the effective current $J^{eff}_{Q,v}$ consists of $Q^{+}$
to $Q^{+}$ and $Q^{-}$ to $Q^{-}$ components as well as mixing ones
of $Q^{+}$ to $Q^{-}$ and $Q^{-}$ to $Q^{+}$. When the effective
field $Q_v$ in $J^{eff}_{Q,v}$ acts on a specific hadron state, the
state will pick up the proper component. That is, the hadron
containing a single heavy quark (antiquark) picks up $Q^{+}$
($Q^{-}$) and cancels $Q^{-}$ ($Q^{+}$).

Now the heavy quark expansion for any heavy-to-heavy transition
matrix elements can be represented as
\begin{eqnarray}
   \label{matrixHQE}
 {\cal A}  &=& \langle H'_{v'} | J_{Q,v} e^{i\int d^4x {\cal L}^{(1/\hat{m}_Q)}_{Q,v}} | H_v\rangle
  =  \langle H'_{v'} | J^{eff}_{Q,v}
        | H_v\rangle           \nonumber\\
  &=& \langle H^{\prime}_{v^{\prime}} | \bar{Q}'_{v'} \Gamma Q_v |
        H_v \rangle
        -\frac{1}{2\hat{m}_Q} \langle H^{\prime}_{v'}| \bar{Q}'_{v'} O_{1}(\Gamma) Q_v | H_v \rangle
        -\frac{1}{2\hat{m}_{Q'}} \langle H^{\prime}_{v'}| \bar{Q}'_{v'} O_{1}^{\prime}(\Gamma)
        Q_v | H_v \rangle  \nonumber \\
        && -\frac{1}{4\hat{m}_Q^2} \langle H^{\prime}_{v'}| \bar{Q}'_{v'} O_{2}(\Gamma)Q_v
        | H_v \rangle
        -\frac{1}{4\hat{m}_{Q'}^2} \langle H^{\prime}_{v'}| \bar{Q}'_{v'} O_{2}^{\prime}(\Gamma)Q_v
        | H_v \rangle  \nonumber \\
        && +\frac{1}{4\hat{m}_Q^2}\langle H^{\prime}_{v'}| \bar{Q}'_{v'} O_{3}(\Gamma)Q_v |
        H_v \rangle
        +\frac{1}{4\hat{m}_{Q'}^2} \langle H^{\prime}_{v'}|\bar{Q}'_{v'} O_{3}^{\prime}(\Gamma)Q_v
       | H_v \rangle  \nonumber \\
       && +\frac{1}{4\hat{m}_{Q'} \hat{m}_Q} \langle H^{\prime}_{v'}|\bar{Q}'_{v'} O_{4}(\Gamma)Q_v |
        H_v \rangle
      +O\Big( \frac{1}{\hat{m}_{Q^{(\prime)}}^3} \Big) .
\end{eqnarray}

When $v^{\prime}=v$, one gets from Eqs.(\ref{normalizationQCD})-(\ref{matrixHHv}) and (\ref{matrixHQE})
 \begin{eqnarray}
    \label{LambdaH}
       \bar{\Lambda}_{H}&=&\bar{\Lambda}-\frac{1}{2\hat{m}_Q} \langle H_v | \bar{Q}_v O_{1}(\VS )
        Q_v | H_v \rangle
       -\frac{1}{4 \hat{m}^2_Q } \langle H_v | \bar{Q}_v (O_{2}(\VS )-O_{3}
         (\VS ) Q_v | H_v \rangle  \nonumber\\
      &&+\frac{1}{8 \hat{m}^2_Q } \langle H_v | \bar{Q}_v O_{4}(\VS )Q_v | H_v \rangle
        +O\Big(\frac{1}{\hat{m}^3_Q }\Big) .
 \end{eqnarray}
It is seen from Eq.(\ref{LambdaH}) that the heavy flavor dependence
of $\bar{\Lambda}_H$ can be attributed to the heavy-to-heavy
transition matrix elements, or relevant wave functions. The heavy
hadron mass is then given by
    \begin{equation}
    \label{eq:48}
      m_H=m_Q +\bar{\Lambda}_H=m_Q +\bar{\Lambda}+ O(1/{\hat{m}_Q})= \hat{m}_Q \left(1 + O(1/{\hat{m}^{2}_Q})\right),
    \end{equation}
which is the fact that the hadron mass consists of the dressed heavy
quark mass and the residual mass suppressed by $1/{\hat{m}_Q}$.

To be concrete, we study the weak transition matrix elements between
ground state pseudoscalar and vector mesons. They can be described
by 18 form factors:
\begin{eqnarray}
  \label{formfactordef}
   &&\hspace{-0.7cm} \langle D(v^{\prime})| \bar{c}\gamma^{\mu} b | B(v)\rangle
     =\sqrt{m_D m_B}[h_{+}(\omega)(v+v^{\prime})^{\mu}+h_{-}(\omega)(v-v^{\prime})^{\mu}] ,\nonumber\\
   &&\hspace{-0.7cm}\langle D^{\ast}(v^{\prime},\epsilon^{\prime})| \bar{c}\gamma^{\mu} b | B(v)\rangle
         = i \sqrt{m_{D^{\ast}} m_B}
         h_{V}(\omega) \epsilon^{\mu \nu \alpha \beta} \epsilon^{\prime \ast}_{\nu}
         v^{\prime}_{\alpha} v_{\beta} ,\nonumber\\
   &&\hspace{-0.7cm}\langle D^{\ast}(v^{\prime},\epsilon^{\prime})| \bar{c}\gamma^{\mu} \gamma^{5}b | B(v)\rangle
         = \sqrt{m_{D^{\ast}} m_B}
         [h_{A_1}(\omega)(1+\omega) \epsilon^{\prime \ast \mu}
       -h_{A_2}(\omega)(\epsilon^{\prime \ast} \cdot v)v^{\mu} \nonumber \\
  &&\hspace{1cm} -h_{A_3}(\omega)(\epsilon^{\prime \ast} \cdot v)v^{\prime\mu}], \nonumber\\
  &&\hspace{-0.7cm}\langle D^{\ast}(v^{\prime},\epsilon^{\prime})|
  \bar{c}\gamma^{\mu}b|
         B^{\ast}(v,\epsilon)\rangle
       =\sqrt{m_{D^{\ast}} m_{B^{\ast}} }
    \{-(\epsilon\cdot \epsilon^{\prime\ast})[h_{1}(\omega)(v+v^{\prime})^{\mu}
       +h_{2}(\omega)(v-v^{\prime})^{\mu}] \nonumber\\
  &&\hspace{1cm} +h_{3}(\omega)(\epsilon^{\prime\ast}\cdot v)\epsilon^{\mu}
 +h_{4}(\omega)(\epsilon\cdot v^{\prime})\epsilon^{\prime\ast\mu}
       -(\epsilon\cdot v^{\prime})(\epsilon^{\prime\ast}\cdot v)
    [h_{5}(\omega)v^{\mu}+h_{6}(\omega)v^{\prime\mu}] \} ,\nonumber \\
&&\hspace{-0.7cm}\langle D^{\ast}(v^{\prime},\epsilon^{\prime})|
\bar{c}\gamma^{\mu} \gamma^{5}b |
        B^{\ast}(v,\epsilon)\rangle
        =i\sqrt{m_{D^{\ast}} m_{B^{\ast}}}
        \{\epsilon^{\mu\nu\alpha\beta}\{\epsilon_{\alpha}\epsilon^{\prime\ast}_{\beta}
        [h_{7}(\omega)(v+v^{\prime})_{\nu}\nonumber\\
   &&\hspace{1cm}+h_{8}(\omega)(v-v^{\prime})_{\nu}]
   +v^{\prime}_{\alpha}v_{\beta}[h_{9}(\omega)(\epsilon^{\prime\ast}\cdot v)
       \epsilon_{\nu}+h_{10}(\omega)(\epsilon\cdot v^{\prime})
        \epsilon^{\prime\ast}_{\nu}] \} \nonumber\\
  &&\hspace{1cm}+\epsilon^{\alpha\beta\gamma\delta}\epsilon_{\alpha}
        \epsilon^{\prime\ast}_{\beta}v_{\gamma}v^{\prime}_{\delta}
       [h_{11}(\omega)v^{\mu}+h_{12}(\omega)v^{\prime\mu} ] \} ,
 \end{eqnarray}
where $\epsilon^{(')\mu}$ is the polarization vector of the vector
meson, and $\omega$ is the product of the four-velocities of heavy
mesons, $\omega=v\cdot v'$.

On the other hand, for such transitions between $(Q^+ \bar{q})$ states one can rewrite Eq.(\ref{matrixHQE}) as
\begin{eqnarray}
  \label{matrixHQEexp}
     {\cal A}
      &=& \langle H^{\prime}_{v^{\prime}}| \QVBP \Big\{ \Gamma
        -\frac{1}{2\hat{m}_Q} \Gamma \frac{-P_+ }{\bar{\Lambda}+iv\cdot D} \Big(  D^2_{\! \bot}
        +\frac{i}{2}\sigma_{\alpha\beta} F^{\alpha\beta} \Big)
        -\frac{1}{2\hat{m}_{Q'}} \Big( \DCLS
        +\frac{i}{2}\sigma_{\alpha\beta}
        F^{\alpha\beta} \Big)  \nonumber \\
&&       \frac{-P'_+}{\bar{\Lambda}-iv^{\prime}\cdot {\stackrel{\leftarrow}{D}}}\Gamma
                -\frac{1}{4\hat{m}^2_Q} \Gamma
        \frac{P_+}{\bar{\Lambda}+iv\cdot D} \Big[ \Big( D^2_{\! \bot}+\frac{i}{2}\sigma_{\alpha\beta}
        F^{\alpha \beta} \Big)( i v\cdot D-\bar{\Lambda})
        -i v_{\alpha} D_{\beta} F^{\alpha\beta}  \nonumber \\
&& +  v_\alpha
        \sigma_{\mu\nu} D^\mu F^{\nu \alpha} \Big]
          -\frac{1}{4\hat{m}_{Q'}^2} \Big[ (-iv^{\prime} \cdot \DL-\bar{\Lambda})
         \Big( \DCLS+\frac{i}{2} \sigma_{\alpha\beta}
         F^{\alpha\beta} \Big)
    +i F^{\alpha\beta} v'_\alpha \DL_{\beta}  \nonumber \\
&&  -F^{\nu \alpha} \DL^\mu
         v'_\alpha \sigma_{\mu \nu} \Big]
        \frac{P'_+}{\bar{\Lambda}-iv^{\prime} \cdot \DL} \Gamma
       +\frac{1}{4\hat{m}^2_Q} \Gamma \frac{P_+}{\bar{\Lambda}+iv\cdot D}
       \Big( D^2_{\! \bot}+ \frac{i}{2} \sigma_{\alpha\beta}
       F^{\alpha\beta} \Big) \frac{P_+}{\bar{\Lambda}+iv\cdot D} \nonumber\\
&&       \Big( D^2_{\! \bot}+ \frac{i}{2} \sigma_{\gamma\delta}
       F^{\gamma\delta} \Big)
     +\frac{1}{4\hat{m}^2_{Q'}} \Big( \DCLS + \frac{i}{2} \sigma_{\alpha\beta}
       F^{\alpha\beta} \Big) \frac{P'_+}{\bar{\Lambda}-iv' \cdot \DL}      \Big( \DCLS + \frac{i}{2} \sigma_{\gamma\delta}
       F^{\gamma\delta} \Big) \nonumber \\
   && \frac{P'_+}{\bar{\Lambda}-iv' \cdot \DL}
       \Gamma
    +\frac{1}{4\hat{m}_Q \hat{m}_{Q'}} \Big( \DCLS + \frac{i}{2} \sigma_{\alpha\beta}
       F^{\alpha\beta} \Big) \frac{P'_+}{\bar{\Lambda}-iv' \cdot \DL}
      \Gamma \frac{P_+}{\bar{\Lambda}+iv\cdot D} \nonumber \\
&&       \Big( D^2_{\! \bot}+ \frac{i}{2} \sigma_{\gamma\delta}
       F^{\gamma\delta} \Big)
    +O\Big( 1/\hat{m}^3_{Q^{(')}} \Big) \Big\} \QV | H_v \rangle
 \end{eqnarray}
with $\sigma^{\alpha \beta}=\frac{i}{2}[\gamma^{\alpha},
\gamma^{\beta}]$, $P'_{+}=\frac{1+{v\hspace{-0.15cm}\slash}^{\prime}}{2}$,
and the gluon field strength tensor $F^{\alpha \beta}=[D^{\beta},D^{\alpha}]$.
Then a set of heavy flavor and spin independent wave functions can be introduced as
follows,
\begin{eqnarray}
    \label{wavefunctiondef}
 && \langle M^{\prime}_{v^{\prime}} | \QVBP\Gamma \QV | M_v \rangle =-\xi(\omega)
     Tr[\bar{\cal M}^{\prime}\Gamma {\cal M}],
   \nonumber   \\
  &&\langle M^{\prime}_{v^{\prime}}| \QVBP\Gamma \frac{-P_+}{\bar{\Lambda}+iv\cdot D}
  D^2_{\! \bot}
   \QV | M_v \rangle =-\kappa_1(\omega) \frac{1}{\bar{\Lambda}} Tr[\bar{\cal M}^{\prime}\Gamma {\cal M}],
\nonumber      \\
 && \langle M^{\prime}_{v^{\prime}}| \QVBP\Gamma \frac{-1}{\bar{\Lambda}+iv\cdot D} P_{+} \frac{i}{2}
      \sigma_{\alpha\beta}F^{\alpha\beta}\QV | M_v \rangle =\frac{1}{\bar{\Lambda}}Tr[ \kappa_{\alpha\beta}(v,v^{\prime})
      \bar{\cal M}^{\prime}\Gamma P_{+}\frac{i}{2}\sigma^{\alpha\beta}{\cal M} ] , \nonumber\\
 && \langle M^{\prime}_{v^{\prime}}| \QVBP\Gamma \frac{P_+}{\bar{\Lambda}+iv\cdot D}
     [\DC^2(i v\cdot D-\bar{\Lambda})-i v_\mu D_\nu F^{\mu\nu}]\QV | M_v\rangle
      =-\varrho_1(\omega)\frac{1}{\bar{\Lambda}}Tr[\bar{\cal M}^{\prime}\Gamma {\cal M}] ,\nonumber \\
 &&  \langle M^{\prime}_{v^{\prime}} | \QVBP\Gamma
   \frac{P_+}{\bar{\Lambda}+iv\cdot D}
      \Big[ \frac{i}{2}\sigma_{\alpha\beta}F^{\alpha\beta}(i v\cdot D-\bar{\Lambda})+
      v_\alpha \sigma_{\mu\nu} D^\mu F^{\nu\alpha} \Big] \QV | M_v \rangle  \nonumber\\
 && \hspace{2cm} =\frac{1}{\bar{\Lambda}}Tr[\varrho_{\alpha\beta}(v,v^{\prime})\bar{\cal M}^{\prime}\Gamma P_{+}
      \frac{i}{2}\sigma^{\alpha\beta}{\cal M}] , \nonumber \\
&&  \langle M^{\prime}_{v^{\prime}}| \QVBP\Gamma \frac{P_+}{\bar{\Lambda}+iv\cdot D}
  D^2_{\! \bot}
  \frac{P_+}{\bar{\Lambda}+iv\cdot D} D^2_{\!\bot} \QV | M_v \rangle
      =-\chi_1(\omega)\frac{1}{\bar{\Lambda}^2} Tr[\bar{\cal M}^{\prime}\Gamma {\cal M}] , \nonumber\\
&& \langle M^{\prime}_{v^{\prime}}| \QVBP\Gamma
  \frac{P_+}{\bar{\Lambda}+iv\cdot D}
 \Big[ D^2_{\! \bot} \frac{P_+}{\bar{\Lambda}+iv\cdot D} \frac{i}{2} \sigma_{\alpha\beta}
 F^{\alpha\beta} + \frac{i}{2}\sigma_{\alpha\beta} F^{\alpha\beta}
\frac{P_+}{\bar{\Lambda}+iv\cdot D} D^2_{\!\bot} \Big] \QV | M_v \rangle \nonumber\\
  &&\hspace{2cm} =\frac{1}{\bar{\Lambda}^2} Tr[\chi_{\alpha\beta}(v,v^{\prime})\bar{\cal M}^{\prime}\Gamma P_{+}
      \frac{i}{2}\sigma^{\alpha\beta}{\cal M}],   \nonumber\\
&&  \langle M^{\prime}_{v^{\prime}} | \QVBP\Gamma \frac{P_+}{\bar{\Lambda}+iv\cdot D}
    \frac{i}{2}\sigma_{\alpha\beta} F^{\alpha\beta} \frac{P_+}{\bar{\Lambda}+iv\cdot D}
    \frac{i}{2}\sigma_{\gamma\delta}F^{\gamma\delta}\QV | M_v \rangle \nonumber\\
  &&\hspace{2cm}  =- \frac{1}{\bar{\Lambda}^2} Tr[\chi_{\alpha\beta\gamma\delta}(v,v^{\prime})
     \bar{\cal M}^{\prime}\Gamma P_{+}
    \frac{i}{2}\sigma^{\alpha\beta} P_{+} \frac{i}{2}\sigma^{\gamma\delta} {\cal M}] ,  \nonumber \\
 &&\langle M^{\prime}_{v^{\prime}} | \QVBP \DCLS \frac{P^{\prime}_{+}}{\bar{\Lambda}-iv' \cdot \DL}
     \Gamma \frac{P_+}{\bar{\Lambda}+iv\cdot D} D^2_{\!\bot} \QV | M_v \rangle
     =-\eta_{1}(\omega) \frac{1}{\bar{\Lambda}^2} Tr[\bar{\cal M}^{\prime}\Gamma {\cal M}] ,\nonumber \\
 && \langle M^{\prime}_{v^{\prime}} | \QVBP  \DCLS
\frac{P^{\prime}_{+} }{\bar{\Lambda} -iv'\cdot \DL}
     \Gamma \frac{P_+}{\bar{\Lambda}+iv\cdot D} \frac{i}{2} \sigma_{\alpha\beta}
     F^{\alpha\beta} \QV | M_v \rangle  \nonumber\\
    &&\hspace{2cm}  =\frac{1}{\bar{\Lambda}^2} Tr[\eta_{\alpha\beta}(v,v^{\prime})
      \bar{\cal M}^{\prime}\Gamma P_{+}\frac{i}{2}\sigma^{\alpha\beta}{\cal M}], \nonumber \\
&&  \langle M^{\prime}_{v^{\prime}} | \QVBP \frac{i}{2} \sigma_{\alpha\beta} F^{\alpha\beta}
      \frac{P'_{+}}{\bar{\Lambda}-iv' \cdot \DL}
     \Gamma \frac{P_+}{\bar{\Lambda}+iv\cdot D} \frac{i}{2} \sigma_{\gamma\delta}
     F^{\gamma\delta} \QV | M_v \rangle   \nonumber \\
&&\hspace{2cm}  =-\frac{1}{\bar{\Lambda}^2} Tr[\eta_{\alpha\beta\gamma\delta}(v,v')
      \bar{\cal M}^{\prime}\frac{i}{2}\sigma^{\alpha\beta} P'_{+} \Gamma P_{+}
      \frac{i}{2}\sigma^{\gamma\delta}{\cal M}]  ,
  \end{eqnarray}
where ${\cal M}$ is the spin wave function
     \begin{eqnarray}
     \label{eq:53}
       {\cal M}(v)=\sqrt{\bar{\Lambda}}P_{+}
         \left\{
           \begin{array}{cl}
              -\gamma^{5} & \mbox{for pseudoscalar meson} \\
              \epsilon\hspace{-0.15cm}\slash & \mbox{for vector meson}
           \end{array}
         \right.
     \end{eqnarray}
and $\bar{\cal M} \equiv \gamma^{0}{\cal M}^{\dagger}\gamma^{0}$.
The decomposition of the tensors $\kappa_{\alpha
\beta}(v,v^{\prime})$, $\varrho_{\alpha \beta}(v,v^{\prime})$,
  $\chi_{\alpha \beta}(v,v^{\prime})$, $\eta_{\alpha \beta}(v,v^{\prime})$,
  $\chi_{\alpha\beta\gamma\delta}(v,v^{\prime})$, and $\eta_{\alpha\beta\gamma\delta}(v,v^{\prime})$ are
  the same as that presented in the Appendix B of Ref.\cite{WYWang1817}.
For simplicity, when the variable $\omega$ is not written
explicitly, we refer to the zero recoil values of relevant
functions, i.e. $h_{A_1}=h_{A_1}(1)$, $\kappa_1=\kappa_1(1)$, etc.

With the definition in Eqs.(\ref{formfactordef}) and
(\ref{wavefunctiondef}), one obtains from Eqs.(\ref{matrixHHv}) and
(\ref{matrixHQEexp})
 \begin{eqnarray}
     \label{Lambdahadronpse}
       \bar{\Lambda}_{D(B)}&=&
          \bar{\Lambda}-\frac{1}{\hat{m}_{c(b)}}
          (\kappa_1+3\kappa_2)
          -\frac{1}{2\hat{m}^2_{c(b)} \bar{\Lambda}}(\varrho_{1} \bar{\Lambda}
          -3\varrho_{2} \bar{\Lambda}-\chi_1-3\chi_2 +3\chi_{4}\nonumber\\
        &&  +9\chi_{5}+6\chi_{6})
          +\frac{1}{4\hat{m}^2_{c(b)} \bar{\Lambda}}(\eta_{1}+6\eta_{2}-3\eta_{4}
          -9\eta_{5}-6\eta_{6})
          +O\Big(\frac{1}{\hat{m}_{c(b)}^3}\Big) , \\
     \label{Lambdahadronvec}
       \bar{\Lambda}_{D^{\ast}(B^{\ast})}&=&
          \bar{\Lambda}-\frac{1}{\hat{m}_{c(b)}} (\kappa_1-\kappa_2)
          -\frac{1}{2\hat{m}^2_{c(b)} \bar{\Lambda}}(\varrho_{1} \bar{\Lambda}
          -\varrho_{2} \bar{\Lambda}-\chi_1+\chi_2 \nonumber \\
         && +3\chi_{4}+\chi_{5}-2\chi_{6})
          +\frac{1}{4\hat{m}^2_{c(b)} \bar{\Lambda}}(\eta_{1}-2\eta_{2}-3\eta_{4}
          -\eta_{5}+2\eta_{6}) +O\Big( \frac{1}{\hat{m}_{c(b)}^3}\Big)  ,
     \end{eqnarray}
where the normalization of the Isgur-Wise function $\xi(1)=1$
\cite{Is.Wi} has been used.

At the zero recoil point, Eqs.(\ref{matrixHHv}) and
(\ref{formfactordef})-(\ref{Lambdahadronvec}) yield
\begin{eqnarray}
    \label{formfactorinwavefunction}
    h_{+}&=&1+\frac{1}{8\bar{\Lambda}^2}\Big( \frac{1}{\hat{m}_b}-\frac{1}{\hat{m}_c}\Big)^2
        \Big[ (\kappa_{1}+3\kappa_{2})^2
         -\eta_{1}-6\eta_{2}+3\eta_{4}+9\eta_{5}+6\eta_{6}\Big] ,\nonumber\\
   h_{A_{1}}&=&1+\frac{1}{8\bar{\Lambda}^2} \Big[ \frac{1}{\hat{m}_b}(\kappa_1+3\kappa_2)
         -\frac{1}{\hat{m}_c}(\kappa_1  -\kappa_2) \Big]^2
       -\frac{1}{8\hat{m}^2_b \bar{\Lambda}^2}
         (\eta_{1}+6\eta_{2}-3\eta_{4}-9\eta_{5}-6\eta_{6}) \nonumber\\
  &&-\frac{1}{8\hat{m}^2_c \bar{\Lambda}^2}
         (\eta_{1}-2\eta_{2} -3\eta_{4} -\eta_{5}+2\eta_{6})
       +\frac{1}{4\hat{m}_b \hat{m}_c \bar{\Lambda}^2}
         (\eta_{1}+2 \eta_{2}+\eta_{4}+3\eta_{5}+2\eta_{6}) , \nonumber\\
     h_{1}&=&1+\frac{1}{8\bar{\Lambda}^2} \Big(\frac{1}{\hat{m}_b}-\frac{1}{\hat{m}_c}\Big)^2
      \Big[(\kappa_{1}-\kappa_{2})^2
       -\eta_1+2\eta_2+3\eta_{4}+\eta_{5}-2\eta_{6} \Big] , \nonumber \\
    h_{7}&=&- \Big [1+\frac{1}{8\bar{\Lambda}^2}\Big(\frac{1}{\hat{m}_b}-\frac{1}{\hat{m}_c}\Big)^2
      (\kappa_{1}-\kappa_{2})^2
       -\frac{1}{8\bar{\Lambda}^2}\Big(\frac{1}{\hat{m}^2_b}+\frac{1}{\hat{m}^2_c}\Big)
       (\eta_1-2\eta_2-3\eta_{4}-\eta_{5}+2\eta_{6})  \nonumber\\
         &&     +\frac{1}{4\hat{m}_b \hat{m}_c
    \bar{\Lambda}^2}(\eta_1-2\eta_2+\eta_{4}-\eta_{5}
     -2\eta_{6}) \Big] .
   \end{eqnarray}
Then it is clear that all matrix elements in (\ref{formfactordef})
are protected from $1/\hat{m}_Q$ order corrections at zero recoil. Furthermore, one
has $h_{-}(\omega)=h_{2}(\omega)=0$ \cite{WYWang1817} because in the new framework of
HQEFT the effective current $J^{eff}_{Q,v}$ contains only terms with even powers of $\DSC$.

\section{$|V_{cb}|$ from Exclusive B Decays}\label{Sectionvcb}

The $B\to D^* (D)l\nu$ differential decay rates are
   \begin{eqnarray}
   \label{widthb2ds}
  \frac{d\Gamma(B\rightarrow D^{\ast}l\nu)}{d\omega} &=&\frac{G^2_F}{48\pi^3}(m_B-m_{D^{\ast}})^2
    m^3_{D^{\ast}}\sqrt{\omega^2-1}(\omega+1)^2 \nonumber\\
    &&  \times  \Big[1+\frac{4\omega}{\omega+1}
    \frac{m^2_B-2\omega m_B m_{D^{\ast}}+m^2_{D^{\ast}}}
    {(m_B-m_{D^{\ast}})^2}\Big] |V_{cb} |^2 {\cal F}^2(\omega) ,\\
\label{widthb2d}
    \frac{d\Gamma(B\rightarrow Dl\nu)}{d\omega}&=&\frac{G^2_F}{48\pi^3}(m_B+m_{D})^2
    m^3_{D}(\omega^2-1)^{3/2} \vert V_{cb} \vert^2 {\cal G}^2(\omega)
   \end{eqnarray}
   with
\begin{eqnarray}
   \label{zerorecoilF}
     {\cal F}(1) &=&
     \eta_{A} h_{A_1}(1) ,\\
   \label{zerorecoilG}
     {\cal G}(1) &=& \eta_{V} \Big[h_{+}(1)-\frac{m_B-m_D}{m_B+m_D}
      h_{-}(1)\Big] ,
\end{eqnarray}
where the QCD radiative corrections to two loops give the short
distance coefficients $ \eta_A=0.960\pm 0.007 $ and $\eta_V=1.022\pm
0.004$ \cite{AC4124}.

The form factors $h_i$ contain long distance effects and can be
estimated by nonperturbative methods such as lattice simulations,
QCD sum rules or quark models. Here we do not perform such
calculations but try to make model independent prediction on
$|V_{cb}|$ using the HQE discussed in the previous sections. Suppose
that the residual momenta of the heavy quarks approximately equal
and the longitudinal residual momenta of dressed heavy quarks be
much smaller than the binding energy, we then have in a good
approximation
   \begin{equation}
    \label{approxeq}
     \frac{1}{i\DSP+\bar{\Lambda}} \sim
     \frac{1}{-i\stackrel{\hspace{-0.1cm}\leftarrow}{\DSP}+\bar{\Lambda}} \sim
     \frac{1}{\bar{\Lambda}}
    \end{equation}
which implies $O_3(\Gamma) \sim O^{\prime}_3(\Gamma) \sim
 O_4(\Gamma)$. Consequently, we arrive at
 the following relations among the wave functions:
   \begin{equation}
   \label{eq:69}
    \chi_1=\eta_{1},  \; \;  \chi_2=2\eta_{2} ,  \; \; \chi_{i}=\eta_{i}\; (i=4,5,6) ,
   \end{equation}
which will be adopted in the following discussions.

Since the contribution of the chromomagnetic moment operator is
generally much smaller than that of the kinetic energy operator, we
neglect operators containing two field strength tensors of gluon but
remain those containing only one. As a result, $\chi_{j}$ and
$\eta_{j}$ for $j=4,5,6$ will be neglected. Thus we get from
Eqs.(\ref{Lambdahadronpse})-(\ref{formfactorinwavefunction})
\begin{eqnarray}
\label{massB}
 \bar{\Lambda}_{D(B)}&=&
   \bar{\Lambda}-\frac{1}{\hat{m}_{c(b)}}
  (\kappa_1+3\kappa_2)
  -\frac{1}{4\hat{m}^2_{c(b)} \bar{\Lambda}}( F_{1}+3F_{2} )
   +O\Big(\frac{1}{\hat{m}_{c(b)}^3}\Big) ,  \\
\label{massBS} \bar{\Lambda}_{D^{\ast}(B^{\ast})}&=&
   \bar{\Lambda}- \frac{1}{\hat{m}_{c(b)}} (\kappa_1-\kappa_2)
   -\frac{1}{4\hat{m}^2_{c(b)} \bar{\Lambda}}( F_1 - F_{2} )
   +O\Big(\frac{1}{\hat{m}_{c(b)}^3}\Big)
\end{eqnarray}
and
\begin{eqnarray}
\label{formfactorha1}
    h_{A_{1}}
 &=&1+\frac{1}{8\bar{\Lambda}^2}\Big[\frac{\kappa_1+3\kappa_2}{\hat{m}_b}
     -\frac{\kappa_1-\kappa_2}{\hat{m}_c}\Big]^2
     -\frac{1}{24 \hat{m}^2_b\bar{\Lambda}^2}
     ( 2\bar{\Lambda} \varrho_1 +6\bar{\Lambda} \varrho_2 -F_1-3F_2)  \nonumber \\
 && -\frac{1}{24\hat{m}_c^2 \bar{\Lambda}^2} (
   2\bar{\Lambda} \varrho_1 -2 \bar{\Lambda} \varrho_2 -F_1 +F_2 )
     +\frac{1}{12 \hat{m}_b \hat{m}_c \bar{\Lambda}^2}
     ( 2\bar{\Lambda} \varrho_1 +2\bar{\Lambda} \varrho_2 -F_1-F_2)  , \\
 \label{formfactorhz}
h_{+}&=&
  1+\frac{1}{8\bar{\Lambda}^2}\Big(\frac{1}{\hat{m}_b}-\frac{1}{\hat{m}_c}\Big)^2
     \Big[(\kappa_1+3\kappa_2)^2-\frac{1}{3}
     (2\bar{\Lambda} \varrho_1 +6\bar{\Lambda} \varrho_2 -F_1-3F_2)
      \Big] , \\
\label{formfactorhf} h_{-}&=& 0 ,
    \end{eqnarray}
where $F_1$ and $F_2$ are defined as
\begin{eqnarray}
   \label{defineF1}
   F_{1}=2\bar{\Lambda}\varrho_{1} -3 \eta_{1} , \hspace{1cm}
   F_{2}=2\bar{\Lambda}\varrho_{2} -6 \eta_{2} .
\end{eqnarray}

\begin{figure}
\centering
\subfigure[]{
\label{fig1a}
\includegraphics[width=8cm,clip=true]{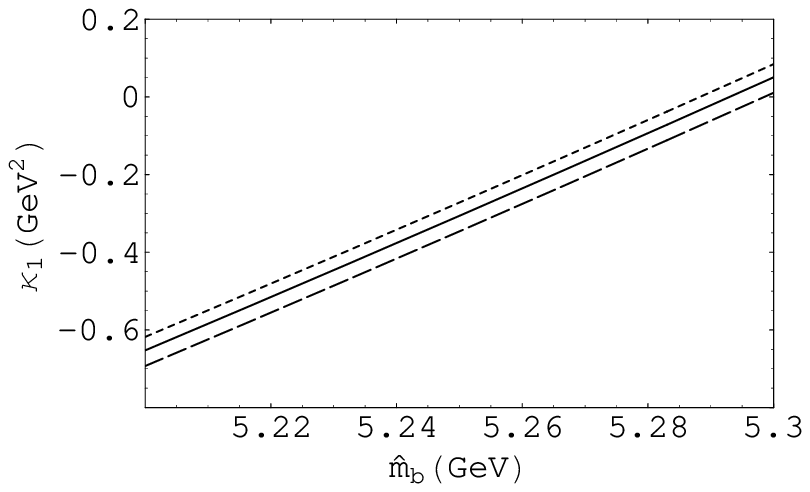}}
\subfigure[]{
\label{fig1b}
\includegraphics[width=8cm,clip=true]{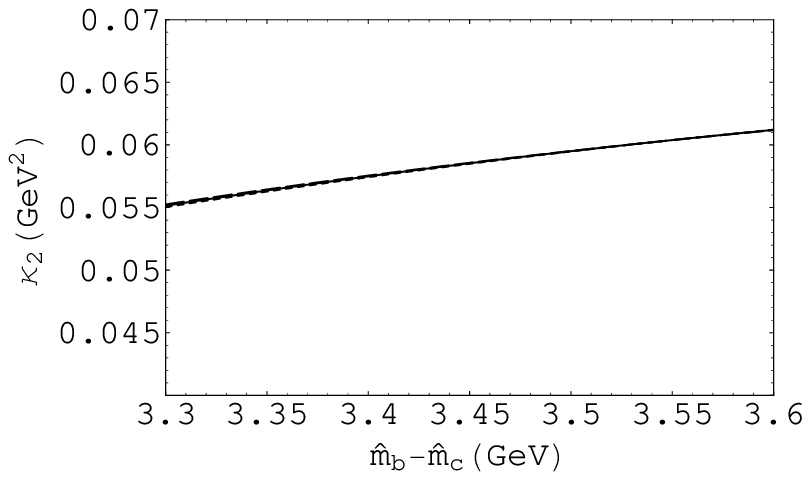}}
\subfigure[]{
\label{fig1c}
\includegraphics[width=8cm,clip=true]{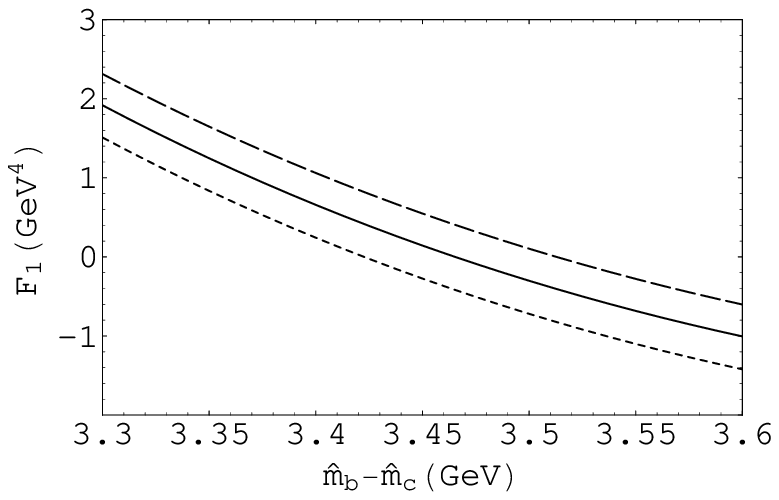}}
\subfigure[]{
\label{fig1d}
\includegraphics[width=8cm,clip=true]{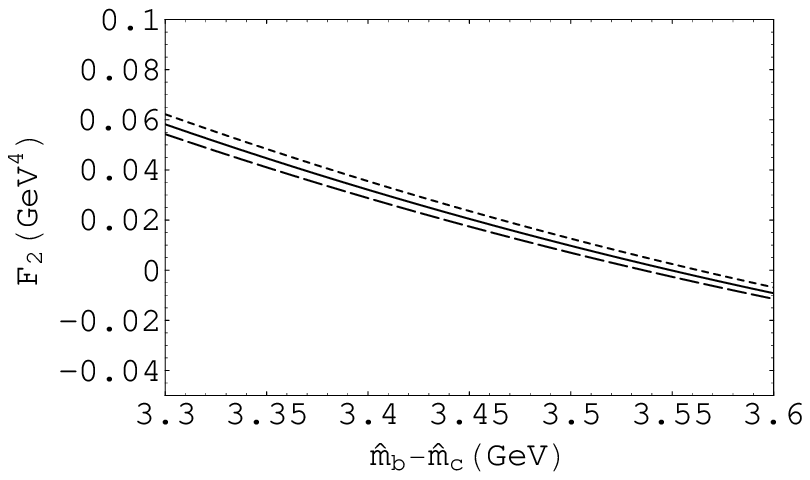}}
\caption{
$\kappa_i$ and $F_i$ as functions
of the variables $\hat{m}_b$ and $\hat{m}_b-\hat{m}_c$. The dashed, solid and dotted curves correspond to
$\hat{m}_b-\hat{m}_c=$3.45, 3.50 and 3.55GeV in (a); and
$\hat{m}_b=$5.23, 5.25 and 5.27GeV in (b)-(d). Figures (c) and (d) are
obtained at $\bar{\Lambda}=0.53$GeV. }
\label{fig1}
\end{figure}

As already mentioned in the previous section, the form factors $h_{A_1}$
and $h_+$ are protected from $1/\hat{m}_Q$ order correction, and
$h_{-}=0$ holds up to order $1/\hat{m}_Q^2$ in our expansion. These
make both the semileptonic decays of $B\to D^* \ell \nu$ and  $B\to D \ell \nu$ the appropriate channels for
the $|V_{cb}|$ extraction. From Eqs.(\ref{massB}) and (\ref{massBS}) one
can estimate the zero recoil values of $\kappa_i$ and $F_i$ from the
bottom and charm meson masses ($m_B=5.279\mbox{GeV}$,
$m_{B^*}=5.325\mbox{GeV}$, $m_D=1.865\mbox{GeV}$ and $m_{D^*}=2.007
\mbox{GeV}$). $\kappa_i$ and $F_i$ as functions of the variables
$\hat{m}_b$ and $\hat{m}_b-\hat{m}_c$ are shown in Fig.\ref{fig1}.
It is found that $\kappa_1$ and $\kappa_2$ are independent of
$\bar{\Lambda}$, and the change of $\bar{\Lambda}$ value only affects
$F_1$ and $F_2$ quite slightly. $\kappa_1$ is sensitive to
$\hat{m}_b$ and also influenced by $\hat{m}_b-\hat{m}_c$. $\kappa_2$
changes slightly against $\hat{m}_b-\hat{m}_c$ but is almost
independent of $\hat{m}_b$. Both $F_1$ and $F_2$ heavily depend on
$\hat{m}_b-\hat{m}_c$, and $F_1$ is also sensitive to $\hat{m}_b$.
When taking
\begin{eqnarray}
\label{variable}
  \hat{m}_b = 5.23 \sim 5.27 \mbox{GeV} , \;\;\; \hat{m}_b-\hat{m}_c = 3.45 \sim 3.55 \mbox{GeV}, \;\;\;
 \bar{\Lambda} = 0.50 \sim 0.56 \mbox{GeV},
\end{eqnarray}
we obtain
\begin{eqnarray}
\label{kappafrommass}
&&  \kappa_1 \approx -0.31 \mbox{GeV$^2$}, \;\;\;
   \kappa_2 \approx 0.06 \mbox{GeV$^2$} ,\nonumber\\
 &&  F_{1} \approx -0.30 \mbox{{GeV}$^4$}  ,\; \;\; \;\;
   F_{2} \approx 0.01 \mbox{{GeV}$^4$} .
 \end{eqnarray}
The data in Eqs.(\ref{variable}) and (\ref{kappafrommass}) are consistent with the results in Ref.\cite{WYWang377}. In that reference
$\kappa_1=-0.50 \pm 0.18 \mbox{GeV}^2$, $\bar{\Lambda}=0.53\pm 0.08\mbox{GeV}$ and
$\kappa_1 \approx -0.43 \mbox{GeV}^2$, $\kappa_2 \approx 0.08 \mbox{GeV}^2$ are given via different methods of analysis on the sum rule equations that include only one-loop perturbative contributions. When the two-loop perturbative contributions are considered in the sum rule for the decay constant, $\kappa_1 \approx -0.34 \mbox{GeV}^2$, $\bar{\Lambda}=0.56\pm 0.08\mbox{GeV}$ are obtained.

\begin{figure}
\centering
\subfigure[]{
\label{fig2a}
\includegraphics[width=8cm,clip=true]{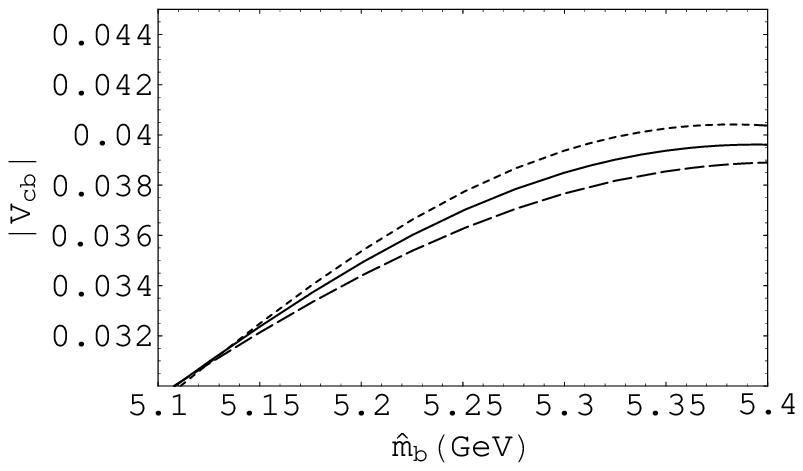}}
\subfigure[]{
\label{fig2b}
\includegraphics[width=8cm,clip=true]{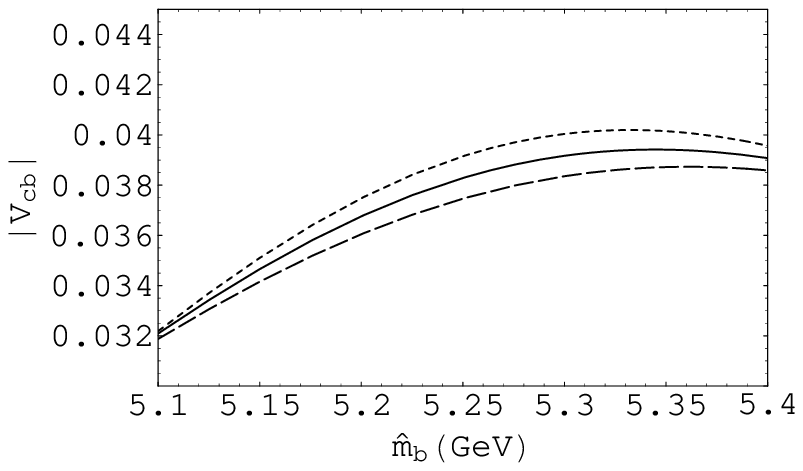}}
\caption{
$|V_{cb}|$ extracted from $B\to D^*\ell \nu$ (a) and $B\to D\ell \nu$ (b).
The dashed, solid and dotted curves correspond to
$\hat{m}_b-\hat{m}_c=$3.4, 3.5 and 3.6GeV, respectively. $\bar{\Lambda}=0.53\mbox{GeV}$ and $\frac{\varrho_i}{\bar{\Lambda}\kappa_i}=1$ are used.}
\label{fig2}
\end{figure}

\begin{figure}
\centering
\subfigure[]{
\label{fig3a}
\includegraphics[width=8cm,clip=true]{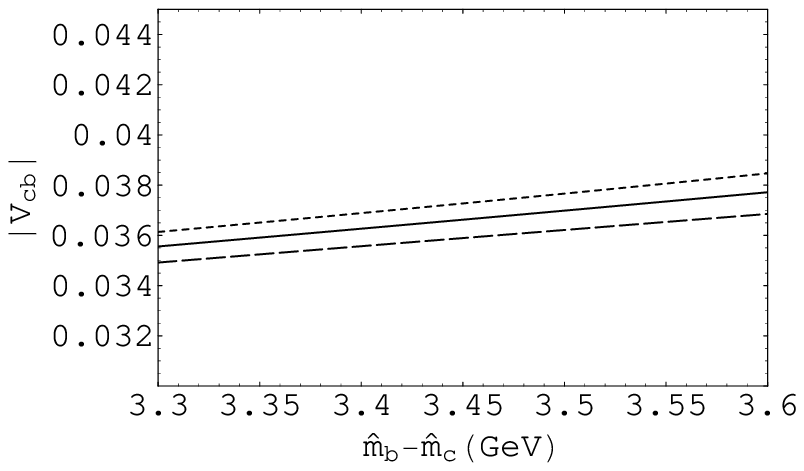}}
\subfigure[]{
\label{fig3b}
\includegraphics[width=8cm,clip=true]{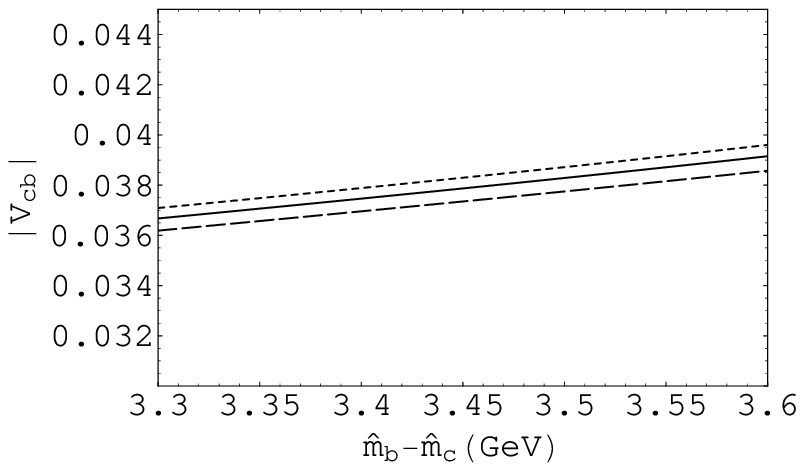}}
\caption{
$|V_{cb}|$ extracted from $B\to D^*\ell \nu$ (a) and $B\to D\ell \nu$ (b).
The dashed, solid and dotted curves correspond to
$\hat{m}_b=$5.23, 5.25 and 5.27GeV, respectively. $\bar{\Lambda}=0.53\mbox{GeV}$ and $\frac{\varrho_i}{\bar{\Lambda}\kappa_i}=1$ are used.}
\label{fig3}
\end{figure}

\begin{figure}
\centering
\subfigure[]{
\label{fig4a}
\includegraphics[width=8cm,clip=true]{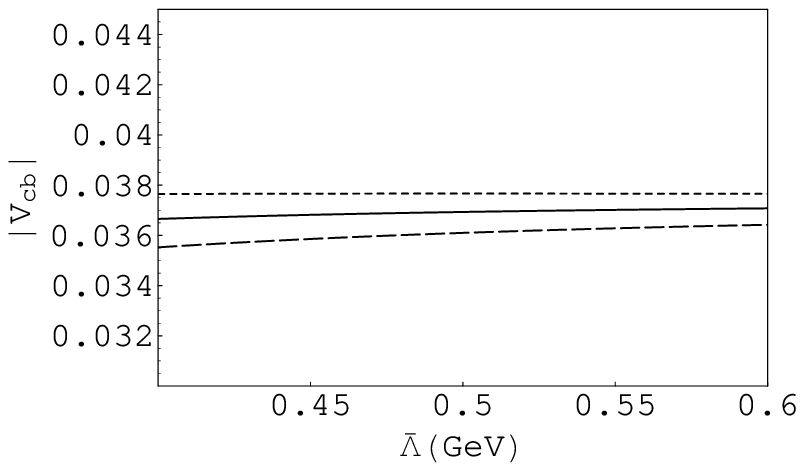}}
\subfigure[]{
\label{fig4b}
\includegraphics[width=8cm,clip=true]{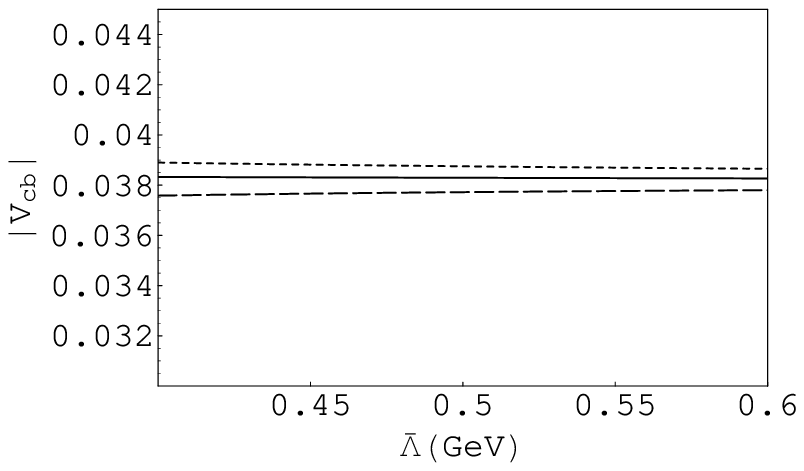}}
\caption{
$|V_{cb}|$ extracted from $B\to D^*\ell \nu$ (a) and $B\to D\ell \nu$ (b).
The dashed, solid and dotted curves correspond to
$\hat{m}_b=$5.23, 5.25 and 5.27GeV, respectively. $\hat{m}_b-\hat{m}_c=3.5\mbox{GeV}$ and $\frac{\varrho_i}{\bar{\Lambda}\kappa_i}=1$ are used.}
\label{fig4}
\end{figure}

\begin{figure}
\centering
\includegraphics[width=8cm,clip=true]{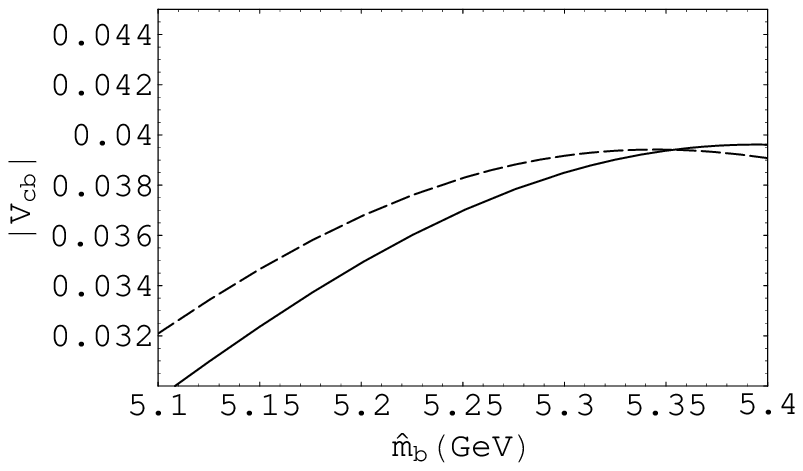}
\caption{
$|V_{cb}|$ extracted from $B\to D^*\ell \nu$ (solid) and $B\to D\ell \nu$ (dashed) decays. $\hat{m}_b-\hat{m}_c=$3.5GeV, $\bar{\Lambda}=0.53\mbox{GeV}$ and $\frac{\varrho_i}{\bar{\Lambda}\kappa_i}=1$ are used.}
\label{fig5}
\end{figure}

$h_{A_1}$ and $h_+$ in (\ref{formfactorha1}) and
(\ref{formfactorhz}) also depend on $\varrho_1$ and $\varrho_2$.
Note that the definition of $\kappa_i$ and $\varrho_i$ in
(\ref{wavefunctiondef}) can be written as
\begin{eqnarray}
&&  \langle M^{\prime}_{v^{\prime}} | \QVBP\Gamma
   \frac{1}{\bar{\Lambda}+iv\cdot D} (i\DSC )^2
 \QV | M_v \rangle  =-\kappa_1(\omega) \frac{1}{\bar{\Lambda}} Tr[\bar{\cal M}^{\prime}\Gamma {\cal M}]
     \nonumber\\
&&\hspace{4cm}
  + \frac{1}{\bar{\Lambda}}Tr[\kappa_{\alpha\beta}(v,v^{\prime})
      \bar{\cal M}^{\prime}\Gamma P_{+}\frac{i}{2}\sigma^{\alpha\beta}{\cal M}] , \\
&& \langle M^{\prime}_{v^{\prime}} | \QVBP\Gamma
  \frac{1}{\bar{\Lambda} + iv\cdot D}
     (i\DSC)( i\DSP + \bar{\Lambda})(i\DSC)   \QV | M_v \rangle
      =-\varrho_1(\omega)\frac{1}{\bar{\Lambda}}Tr[\bar{\cal M}^{\prime}\Gamma {\cal M}]
      \nonumber\\
&&\hspace{4cm}
   +\frac{1}{\bar{\Lambda}}Tr[\varrho_{\alpha\beta}(v,v^{\prime})\bar{\cal M}^{\prime}\Gamma P_{+}
      \frac{i}{2}\sigma^{\alpha\beta}{\cal M}] .
  \end{eqnarray}
Then the approximation (\ref{approxeq}) implies
$\frac{\varrho_i}{\bar{\Lambda}\kappa_i} \approx 1$.
The resulting $|V_{cb}|$ value is shown in Figs.\ref{fig2}-\ref{fig5}.
Using
(\ref{variable}) and allowing $\varrho_i$ change in the range
\begin{equation}
\label{rhokappa}
\frac{\varrho_i}{\bar{\Lambda} \kappa_i} = 0\sim 2 \;\;\; (i=1,2) ,
\end{equation}
we get
\begin{eqnarray}
\label{hA1B}
h_{A_1}(1) &=& 1.014 \pm 0.034 , \\
\label{hzB}
h_+(1) &=& 0.997 \pm 0.025 .
\end{eqnarray}

Consequently, the averages of measurements \cite{PDGCA667}
     \begin{eqnarray}
     \label{exp1}
     \vert V_{cb}\vert {\cal F}(1)&=&0.0360 \pm 0.0013,\\
     \label{exp2}
     \vert V_{cb}\vert {\cal G}(1)&=&0.039 \pm 0.004
     \end{eqnarray}
give
\begin{eqnarray}
  \label{vcbDsB}
  |V_{cb} |_{B\to D^*} &=& 0.0370 \pm 0.0013_{\mbox{exp}}
     \pm  0.0015_{\mbox{th}}   ,\\
  \label{vcbDB}
  | V_{cb} |_{B\to D} &=& 0.0383  \pm 0.0039_{\mbox{exp}}
     \pm 0.0011_{\mbox{th}}  .
\end{eqnarray}
So the $|V_{cb}|$ values extracted from $B\to D^*\ell
\nu$ and $B\to D\ell \nu$ decays are consistent within the errors of
experimental data. It is noticed that the value extracted from $B\to
D \ell \nu$ suffers from relatively larger experimental uncertainty,
which can be seen in Eq.(\ref{exp2}).
The result for $|V_{cb}|$ in (\ref{vcbDsB}) is marginally consistent with the
value given in Ref.\cite{PDGCA667} ($0.0386\pm 0.0013$) but has a smaller center value. For more precise determination of $|V_{cb}|$, it would be helpful to evaluate $1/\hat{m}^2_Q$ order wave functions such as $F_i$ and $\varrho_i$ through other methods like lattice or QCD sum rule calculation.

\section{Conclusions}\label{Sectionsummary}

We have briefly reviewed the derivation of a heavy quark effective
field theory. This HQEFT is complete in that the effective Lagrangian contains the heavy quark-antiquark coupling terms, which appear as finite mass corrections. Unlike the usual naive
heavy quark-hadron duality, we do not simply treat the light
components of hadrons as spectators. Instead, the flavor independent
nonperturbative effects of light degrees of freedom are attributed
to the dressed heavy quark with the dressed mass $\hat{m}_Q=m_Q +
\bar{\Lambda}$ and the total momentum $p_Q=\hat{m}_Q v+\tilde{k}$,
and it is such a dressed heavy quark that dualizes the heavy hadron.
Consequently, HQEFT has been extended into the formulation in terms
of $1/\hat{m}_Q$ expansion. Such an expansion is consistent with the
picture of dressed heavy quark and becomes more convergent. The HQE
of heavy-to-heavy transition matrix elements has been consistently
extended into a $1/\hat{m}_Q$ expansion form, in which the contribution of heavy antiquark (or quark) field is integrated into the effective current.

$|V_{cb}|$ extraction from $B\to D^*(D)\ell \nu$ decays has been
discussed by using the HQE in $1/\hat{m}_Q$. Due to the appropriate
definition of effective states, zero recoil values of the relevant
form factors can be estimated from the hadron masses. Using some approximate relations between wave functions $|V_{cb}|$ is found to be
$ 0.0370 \pm 0.0013_{\mbox{exp}}  \pm
0.0015_{\mbox{th}} $ from $B\to D^* \ell \nu$ decay and
$ 0.0383 \pm 0.0039_{\mbox{exp}}  \pm
0.0011_{\mbox{th}} $ from $B\to D \ell \nu$ decay.
For these two channels, experimental study of $B\to D \ell \nu$ is more difficult and it has larger uncertainties. Nevertheless, the current data of $|V_{cb}|{\cal
G}(1)$ and the form factors extracted within the framework of HQEFT
give the $|V_{cb}|$ value consistent with that from
$B\to D^*\ell \nu$ decay, which shows the reliability of the $1/\hat{m}_Q$ expansion in this application.

%===================================================

\acknowledgments

This work was supported in part by the National Science Foundation
of China (NSFC) under Grant \#No. 10821504, 10805005, 10975170 and
the Project of Knowledge Innovation Program (PKIP) of the Chinese
Academy of Science.

\appendix

\section{}

This appendix is devoted to the derivation of the effective current
$J^{eff}_{Q,v}$ in Eqs.(\ref{neweffectivecurrent}) and (\ref{matrixHQE}).
In particular, we would like to show why one can use (\ref{propagator}) as the propagator of $Q_v$ field.

Firstly, one may notice that the gluon couplings arising from $\DSP$ can be trivialized by the Wilson-line transformation \cite{FHuss295}. One can introduce new field variable $Q^0_v$ by \cite{YLWu5743}
\begin{eqnarray}
Q_v&=&{\cal P} e^{ig \int^{v\cdot x}_{-\infty} d\tau v\cdot A^a T^a} Q^0_v \equiv W(x,v)Q^0_v, \\
\bar{Q}_v&=&\bar{Q}^0_v {\cal P} e^{-ig \int^{v\cdot x}_{-\infty} d\tau v\cdot A^a T^a} \equiv \bar{Q}^0_v W^{-1}(x,v),
\end{eqnarray}
where ${\cal P}$ denotes path ordering with $x^\mu=v^\mu \tau$.
Since
\begin{eqnarray}
 v\cdot D Q_v &=& {\cal P} e^{ig\int^{v\cdot x}_{-\infty} d\tau v\cdot A^a T^a}
 v\cdot \partial Q^0_v ,\\
(\DS - \VS v\cdot D) Q_v&=&{\cal P} e^{ig\int^{v\cdot x}_{-\infty} d\tau v\cdot A^a T^a}
   (\DS-\VS v\cdot D) Q^0_v,
\end{eqnarray}
one can write Eqs.(\ref{JQvexp0})-(\ref{LQvexp2}) as
\begin{eqnarray}
\label{JQ0vexp0}
J_{Q,v}^{(0)}(x) & = &
\sum _{\sigma, \sigma ^{\prime} = \pm 1}e^{i
(\sigma ^{\prime} \hat{m}_{Q'}v' - \sigma \hat{m}_{Q}v)\cdot x}
\bar{Q}^{0'}_{v'}(x)\frac{1+\sigma ^{\prime} \vslash ' }{2}
 W^{-1}(x,v')\Gamma W(x,v) \nonumber \\
&& \frac{1+\sigma \vslash }{2}Q^0_v(x), \\
\label{JQ0vexpcor}
 J_{Q,v}^{(1/\hat{m}_Q)}(x)& = &
\sum _{\sigma, \sigma ^{\prime} = \pm 1}e^{i (\sigma ^{\prime}
\hat{m}_{Q'}v' - \sigma \hat{m}_{Q}v)\cdot x}
\bar{Q}^{0'}_{v'}(x)\frac{1+\sigma ^{\prime} \vslash ' }{2}
\Big [ \frac{1}{2\hat{m}_Q}  W^{-1}(x,v')\Gamma W(x,v) i\DSC \nonumber\\
&&+\frac{1}{2\hat{m}_{Q'}}(-i\overleftarrow\DSC) W^{-1}(x,v')\Gamma W(x,v)
+\frac{1}{4\hat{m}_{Q}^{2}} W^{-1}(x,v')\Gamma W(x,v) \nonumber \\
&&  (i \VS v\cdot \partial +\bar{\Lambda})i\DSC
+\frac{1}{4\hat{m}_{Q'}^{2}}(-i\overleftarrow \DSC)(-i \VS v \cdot \overleftarrow\partial +\bar{\Lambda}') W^{-1}(x,v')\Gamma W(x,v) \nonumber\\
&&+ \frac{1}{4\hat{m}_{Q}\hat{m}_{Q'}}(-i\overleftarrow \DSC) W^{-1}(x,v')\Gamma W(x,v)
  (i\DSC)  \nonumber \\
&&  + O\Big ( \frac{1}{\hat{m}_{Q^{(\prime )}}^{3}}\Big)
    \Big ]
\frac{1+\sigma \vslash }{2}Q^0_v(x)
\end{eqnarray}
and
\begin{eqnarray}
\label{LQ0vexp1}
{\cal L}^{I(1/\hat{m}_Q)}_{Q,v} & = &
 \sum _{\varepsilon = \pm 1} \bar{Q}^0_v \frac{1+ \varepsilon \vslash }{2}
\Big[\frac{(i\DSC)^{2}}{2\hat{m}_Q}+ \frac{1}{4\hat{m}_{Q}^{2}}i\DSC
(i\VS v\cdot \partial +\bar{\Lambda})i\DSC  \nonumber\\
&&+ O\Big(
\frac{1}{\hat{m}_Q^3} \Big) \Big]
\frac{1+ \varepsilon \vslash }{2}  Q^0_v  , \\
\label{LQ0vexp2}
{\cal L}^{II(1/\hat{m}_Q)}_{Q,v} & = & \sum _{\varepsilon = \pm 1}
e^{2i\hat{m}_{Q}v\cdot x\varepsilon }\bar{Q}^0_v \frac{1+ \varepsilon
\vslash }{2}\Big[ \frac{1}{2\hat{m}_Q}(-i\VS v\cdot \overleftarrow\partial  +\bar{\Lambda })i\DSC \nonumber \\
&& +\frac{1}{4\hat{m}_{Q}^{2}}(-i \VS v\cdot \overleftarrow\partial +\bar{\Lambda })
(i\VS v\cdot \partial +\bar{\Lambda})i\DSC \nonumber \\
&&+\frac{1}{4\hat{m}_{Q}^{2}}(-i\overleftarrow \DSC)^{2}i\DSC
  + O\Big(\frac{1}{\hat{m}_Q^3} \Big)
\Big]\frac{1- \varepsilon \vslash }{2} Q^0_v  .
\end{eqnarray}

In terms of $Q^0_v$, the effective Lagrangian in heavy quark limit turns into
\begin{eqnarray}
 \label{LQ0v0}
 {\cal L}^{(0)}_{Q,v}&=& \bar{Q}^0_v (i\VS v\cdot \partial+\bar{\Lambda}) Q^0_v ,
\end{eqnarray}
and the contraction of $Q^0_v$ fields yields the propagator:
\begin{eqnarray}
 \label{propagatorQ0v}
\underbracket{Q^0_v(x),\hspace{0.3cm} \bar{Q}^0_v(y)}&=&\int \frac{d^4k}{(2\pi)^4} e^{-ik\cdot (x-y)} \frac{i}{\VS v\cdot k+\bar{\Lambda}} \nonumber \\
&=&\int \frac{d^4k}{(2\pi)^4} e^{-ik\cdot (x-y)} \frac{i}{\bar{\Lambda}} \sum ^{\infty}_{n=0}\Big(- \VS\frac{v\cdot k}{\bar{\Lambda}}\Big)^n  .
\end{eqnarray}

To illustrate the details of deriving the effective current $J^{eff}_{Q,v}$, we consider as an example the two-point correlation function
\begin{eqnarray}
\label{2pointfunction}
&&\int d^4y T\{\bar{Q}_{v'}(x) \hat{O}_1(\Gamma_1)_{(x)}Q_v(x),\bar{Q}_v(y) \hat{O}_2(\Gamma_2)_{(y)}Q_{v''}(y) \}  \\
 && \hspace{0.5cm}= \int d^4y T\{
\bar{Q}^0_{v'}(x) \hat{O}^0_1(W^{-1}(x,v')\Gamma_1 W(x,v))_{(x)}Q^0_v(x),\nonumber \\
&& \hspace{2cm} \bar{Q}^0_v(y) \hat{O}^0_2(W^{-1}(y,v)\Gamma_2 W(y,v''))_{(y)}Q^0_{v''}(y)
 \},
\end{eqnarray}
where $\hat{O}_1(\Gamma)$ and $\hat{O}_2(\Gamma)$ can be any local operators that may contain $\overleftarrow\DSP$ ($\DSP$) and $\overleftarrow\DSC$ ($\DSC$) on the left (right) of the Dirac matrixes $\Gamma$. $\hat{O}^0_1(\Gamma)$ and $\hat{O}^0_2(\Gamma)$ are obtained from
$\hat{O}_1(\Gamma)$ and $\hat{O}_2(\Gamma)$ by replacing $\overleftarrow\DSP$ and $\DSP$ in the operators with $\VS v\cdot \overleftarrow\partial$ and $\VS v\cdot \partial$, respectively.

Using the propagator (\ref{propagatorQ0v}) for field contraction and applying the integration by parts, the two-point function (\ref{2pointfunction}) becomes
\begin{eqnarray}
&&\int d^4y \bar{Q}^0_{v'}(x) \hat{O}^0_1(W^{-1}(x,v')\Gamma_1 W(x,v))_{(x)} \int \frac{d^4k}{(2\pi)^4} \frac{i}{\bar{\Lambda}}
\sum ^{\infty}_{n=0} \Big[\Big( \frac{i}{\bar{\Lambda}} \VS v\cdot \partial_{(y)} \Big)^n e^{-ik\cdot (x-y)} \Big] \nonumber\\ && \hat{O}^0_2(W^{-1}(y,v)\Gamma_2 W(y,v''))_{(y)} Q^0_{v''}(y)  \nonumber\\
&=&
\int d^4y \bar{Q}^0_{v'}(x) \hat{O}^0_1(W^{-1}(x,v')\Gamma_1 W(x,v))_{(x)} \int \frac{d^4k}{(2\pi)^4} e^{-ik\cdot (x-y)} \frac{i}{\bar{\Lambda}} \sum ^{\infty}_{n=0} \Big( -\frac{i}{\bar{\Lambda}} \VS v\cdot \partial_{(y)} \Big)^n \nonumber\\
&&  \hat{O}^0_2(W^{-1}(y,v)\Gamma_2 W(y,v''))_{(y)}Q^0_{v''}(y)  \nonumber\\
&=&
 \bar{Q}^0_{v'}(x) \hat{O}^0_1(W^{-1}(x,v')\Gamma_1 W(x,v))_{(x)}  \frac{i}{\bar{\Lambda}} \sum ^{\infty}_{n=0} \Big( -\frac{i}{\bar{\Lambda}} \VS v\cdot \partial_{(x)} \Big)^n \nonumber\\
&&  \hat{O}^0_2(W^{-1}(x,v)\Gamma_2 W(x,v''))_{(x)}Q^0_{v''}(x)  \nonumber\\
&=&
 \bar{Q}_{v'}(x) \hat{O}_1(\Gamma_1)_{(x)}  \frac{i}{\bar{\Lambda}} \sum ^{\infty}_{n=0} \Big( -\frac{i}{\bar{\Lambda}} {\DSP}_{(x)} \Big)^n
  \hat{O}_2(\Gamma_2 )_{(x)} Q_{v''}(x)  \nonumber \\
&=&  \bar{Q}_{v'}(x) \hat{O}_1(\Gamma_1)_{(x)} \frac{i}{i{\DSP}_{(x)} +\bar{\Lambda}} \hat{O}_2(\Gamma_2)_{(x)}Q_{v''}(x) ,
\end{eqnarray}
the final expression of which can be obtained directly from (\ref{2pointfunction}) with using (\ref{propagator}) as $Q_v$ field propagator.

With the same techniques it is then easy to derive Eq.(\ref{neweffectivecurrent}) from (\ref{JQvexp0})-(\ref{LQvexp2}).

\newpage

\begin{thebibliography}{99}

\bibitem{symmetry} E. V. Shuryak, Phys. Lett. {\bf B 93}, 134 (1980);
    Nucl. Phys. {\bf B 198}, 83 (1982).
\bibitem{symmetry2} S. Nussinov and W. Wetzel, Phys. Rev. {\bf D 36}, 130 (1987).
\bibitem{Is.Wi} N. Isgur and M. Wise, Phys. Lett. {\bf B 232}, 113 (1989);
   {\bf B 237}, 527 (1990); {\bf B 206}, 681 (1988).

\bibitem{MBVoloshin292} M. B. Voloshin and M. A. Shifman, Sov. J. Nucl. Phys. {\bf 45}, 292 (1987); {\bf 47},
    199 (1988).

\bibitem{HGeorgi447} H. Georgi, Phys. Lett. {\bf B 240}, 447 (1990).

\bibitem{TMannel2388} T. Mannel and Z. Ryzak, Phys. Lett. {\bf B 247}, 2388 (1990).

\bibitem{MELuke447} M. E. Luke, Phys. Lett. {\bf B 252}, 447 (1990).

\bibitem{BGrinstein253} B. Grinstein, Nucl. Phys. {\bf B 339}, 253 (1990).

\bibitem{AFFalk1} A. Falk, H. Georgi, B. Grinstein and M. B. Wise, Nucl. Phys. {\bf B 343}, 1 (1990).

\bibitem{AFFalk185} A. F. Falk, B. Grinstein and M. E. Luke, Nucl. Phys. {\bf B 357}, 185 (1991).

\bibitem{TMannel204} T. Mannel, W. Roberts and Z. Ryzak, Nucl. Phys. {\bf B 368}, 204 (1992).

\bibitem{BGrinstein34} B. Grinstein, SSCL-Preprint-34, 1992.

\bibitem{TMannel428} T. Mannel, Phys. Rev. {\bf D 50}, 428 (1994).

\bibitem{TMannel396} T. Mannel, Nucl. Phys. {\bf B 413}, 396 (1994).

\bibitem{MNeubert259} M. Neubert, Phys. Rept. {\bf 245}, 259 (1994).

\bibitem{YLWu819} Y. L. Wu, Mod. Phys. Lett. {\bf A 8}, 819 (1993).

\bibitem{WYWang1817} W. Y. Wang, Y. L. Wu and Y. A. Yan, Int. J. Mod. Phys. {\bf A 15}, 1817 (2000).

\bibitem{YLWu1303} Y. L. Wu, Y. A. Yan, M. Zhong, Y. B. Zuo and W. Y. Wang,
Mod. Phys. Lett. {\bf A 18}, 1303 (2003).

\bibitem{YLWu5743} Y. L. Wu, Int. J. Mod. Phys. {\bf A 21}, 5743 (2006); and references therein.

\bibitem{YAYan2735} Y. A. Yan, Y. L. Wu and W. Y. Wang, Int. J. Mod. Phys. {\bf A 15}, 2735 (2000).

\bibitem{YLWu285} Y. L. Wu and Y. A. Yan, Int. J. Mod. Phys. {\bf A 16}, 285 (2001).

\bibitem{YBZuo3685} Y. B. Zuo, Y. A. Yan, Y. L Wu and W. Y. Wang, Int. J. Mod. Phys. {\bf A 19}, 3685 (2004).

\bibitem{WYWang1379} W. Y. Wang, Y. L. Wu, Y. A. Yan, M. Zhong and Y. B. Zuo,
Mod. Phys. Lett. {\bf A 19}, 1379 (2004).

\bibitem{WYWang377} W. Y. Wang and Y. L. Wu, Int. J. Mod. Phys. {\bf A 16}, 377 (2001).

\bibitem{WYWang57} W. Y. Wang and Y. L. Wu, Phys. Lett. {\bf B 515}, 57 (2001).

\bibitem{WYWang219} W. Y. Wang and Y. L. Wu, Phys. Lett. {\bf B 519}, 219 (2001).

\bibitem{WYWang2743} W. Y. Wang and Y. L. Wu, M. Zhong, J. Phys. {\bf G 29}, 2743 (2003).


\bibitem{WYWang014024} W. Y. Wang, Y. L. Wu and M. Zhong, Phys. Rev. {\bf D 67}, 014024 (2003).

\bibitem{WYWang228} W. Y. Wang, Y. L. Wu and M. Zhong, Phys. Lett. {\bf B 628}, 228 (2005).

\bibitem{AC4124} A. Czarnecki, Phys. Rev. Lett. {\bf 76}, 4124 (1996).

\bibitem{PDGCA667} Particle Data Group, C. Amsler \textit{et al.}, Phys. Lett. {\bf B 667}, 1 (2008).

\bibitem{FHuss295} F. Hussain, J. G. K\"{o}rner, K. Schilcher, G. Thompson and Y. L. Wu, Phys. Lett. {\bf B 249}, 295 (1990).

\end{thebibliography}
\end{document}